\documentclass[12pt]{extarticle}
\pdfoutput=1 
\usepackage{slashed}
\usepackage{latexsym}
\usepackage{amssymb}
\usepackage{amsmath}
\usepackage{autobreak}
\usepackage{graphicx}
\usepackage{arydshln}
\usepackage{adjustbox}
\usepackage{easybmat}
\usepackage{bbm}
\usepackage{cite}
\usepackage{bm}
\usepackage{adjustbox}
\usepackage{booktabs}
\usepackage{multirow} 
\usepackage{rotating}
\usepackage{mathtools}
\usepackage{lscape}
\usepackage{hyperref}
\usepackage{bm}
\usepackage{color}
\usepackage{fancybox,framed}
\usepackage{lscape}
\usepackage{multirow}
\usepackage{fancyhdr}
\usepackage{enumitem}
\usepackage{calrsfs}

\usepackage[most]{tcolorbox} 

\newcommand{\sumintB}[1]{{\hbox{$\sum$}\!\!\!\!\!\!\!\int\,}_{\!\!\!\!\!\!\raise-0.2ex\hbox{$\scriptstyle{#1}$}}}
\newcommand{\sumintF}[1]{{\hbox{$\sum$}\!\!\!\!\!\!\!\int\,}_{\!\!\!\!\!\!\raise-0.2ex\hbox{$\scriptstyle{\{#1\}}$}}}

\definecolor{block-gray}{gray}{0.95}

\definecolor{shadecolor}{rgb}{0.01,0.199,0.1}

\newcommand{\nn }{\nonumber        }
\newcommand{\mc}[1]{\mathcal{#1}}

\newcommand{\blue}[1]{\textcolor{blue}{#1}}

\newtcolorbox{codeSyntax}{
    enhanced,
    frame hidden,
    colback=block-gray,
    boxrule=0pt,
    borderline west={2pt}{0pt}{gray!80!black}
}

\allowdisplaybreaks
\begin{document}
\thispagestyle{empty}
\begin{flushright}
\end{flushright}
\vspace{0.8cm}

\begin{center}
{\Large\sc Phase Transitions in Dimensional\\[0.3cm] Reduction up to Three Loops}
\vspace{0.8cm}

\textbf{
Mikael Chala, Luis Gil and Zhe Ren
}\\
\vspace{1.cm}
{\em {Departamento de F\'isica Te\'orica y del Cosmos,
Universidad de Granada, Campus de Fuentenueva, E--18071 Granada, Spain}}
\vspace{0.5cm}

\texttt{mikael.chala@ugr.es}, \texttt{lgil@ugr.es}, \texttt{zheren@ugr.es}
\end{center}

\begin{abstract}
    We perform the first computation of phase-transition parameters to cubic order in $\lambda\sim m^2/T^2$, where $m$ is the scalar mass and $T$ is the temperature, in a simple model resembling the Higgs sector of the SMEFT. We use dimensional reduction, including 1-loop matching corrections for terms of dimension 6 (in 4-dimensional units), 2-loop contributions for dimension-4 ones and 3-loops for the squared mass. We precisely quantify the size of the different corrections, including renormalisation-group running as well as quantum effects from light fields in the effective theory provided by the Coleman-Weinberg potential, and discuss briefly the implications for gravitational waves. Our results suggest that, for strong phase transitions, 1-loop corrections from dimension-6 operators can compete with 2-loop ones from quartic couplings, but largely surpass those from 3-loop thermal masses.
\end{abstract}

\newpage

\tableofcontents

\section{Introduction}
The dimensional reduction (DR) formalism~\cite{Ginsparg:1980ef,Appelquist:1981vg} has become a cornerstone in the study of equilibrium phenomena in quantum field theory at finite temperature ($T$). By leveraging the hierarchy of scales between the Matsubara modes~\cite{Matsubara:1955ws}, with masses $m_n \sim \pi n T$, and the light field masses, $m \ll T$, DR recasts the dynamics of a 4-dimensional (4D) theory into a simpler static 3-dimensional (3D) effective field theory (EFT)~\cite{Braaten:1995cm,Kajantie:1995dw}, resulting from integrating out the non-zero Matsubara modes, whose effects are encoded in the Wilson coefficients (WC) of local operators.
This framework offers several advantages over direct 4D methods, including the possibility of simulating long-distance non-perturbative physics on the lattice~\cite{Farakos:1994xh,Kajantie:1995kf,Laine:1995np,Laine:1997dy,Gurtler:1997hr,Rummukainen:1998as,Laine:1998jb,Moore:2000jw,Arnold:2001ir,Sun:2002cc,DOnofrio:2015gop,Gould:2022ran}, as well as improved convergence of perturbative expansions~\cite{Croon:2020cgk,Gould:2021oba,Gould:2023ovu}. For these reasons, DR has been widely employed in studies of hot QCD~\cite{Braaten:1994na,Braaten:1995jr,Kajantie:1997tt,Laine:2018lgj,Laine:2019uua,Ghiglieri:2021bom} and, more recently, in the characterisation of phase transitions (PT) beyond the Standard Model (SM)~\cite{Brauner:2016fla,
Andersen:2017ika,Niemi:2018asa,Gorda:2018hvi,Kainulainen:2019kyp,Croon:2020cgk,Gould:2019qek,Niemi:2020hto,Gould:2021ccf,Gould:2021dzl,Schicho:2021gca,Niemi:2021qvp,Camargo-Molina:2021zgz,Niemi:2022bjg,Ekstedt:2022ceo,Gould:2022ran,Ekstedt:2022zro,Biondini:2022ggt,Schicho:2022wty,Lofgren:2021ogg,Gould:2023jbz,Kierkla:2023von,Aarts:2023vsf,Niemi:2024axp,Chala:2024xll,Qin:2024idc,Gould:2024jjt,Camargo-Molina:2024sde,Niemi:2024vzw,Kierkla:2025qyz}. This effort is largely motivated by the prospect of observing stochastic gravitational waves (GW) from PTs in the early universe~\cite{Harry:2006fi,Kawamura:2006up,Ruan:2018tsw,LIGOScientific:2014pky,Caprini:2019egz}, a promising probe of new physics complementary to collider experiments.

While most existing analyses focus on leading-order contributions in the 3-dimensional EFT, typically dominated by dimension-4 interactions, there is mounting evidence that higher-dimensional operators play a decisive role in the dynamics of very strong PTs~\cite{Chala:2024xll,Bernardo:2025vkz,Chala:2025aiz}. (Note that we quote energy dimensions in 4-dimensional units.) In Ref.~\cite{Chala:2024xll}, it was first proven, within a simple model consisting of a real scalar coupled to a fermion~\cite{Gould:2023jbz}, that dimension-6 operators generated at 1 loop through matching can modify the amplitude and peak frequency of GW spectra by orders of magnitude. This result has been further strengthened in Refs.~\cite{Bernardo:2025vkz} and \cite{Chala:2025aiz} in the context of the Abelian Higgs model and the SMEFT~\cite{Buchmuller:1985jz,Grzadkowski:2010es}, respectively, where it was also proven that higher-dimensional operators are relevant to obtain gauge-independent physical results.

However, using standard power counting rules~\cite{Croon:2020cgk}, 1-loop dimension-6 matching corrections are parametrically of the same order as 3-loop thermal masses and 2-loop quartic couplings. Even though dimension-6 operators can be expected to become more relevant at larger values of $m/T$, where PTs occur, a quantitative assessment of the relative size of these corrections remains important. We address this quantitative analysis in the present work. 
To this end, we consider a toy version of the SMEFT, consisting of a complex scalar $\phi$ together with left- and right-handed fermions, including $\phi^6$ corrections that generate a potential barrier at tree level. All sum-integrals arising in the matching up to 3 loops are known. We also briefly comment on the relative importance of different loop corrections in a model with a real scalar and a fermion in which the barrier is provided by a trilinear coupling~\cite{Gould:2023jbz,Chala:2024xll}, but in which 3-loop sum-integrals are not known.

The article is organized as follows. We introduce the main model in section~\ref{sec:theory}, with a discussion of the matching, cancellation of UV and IR divergences and renormalisation-scale dependence. We explore the impact of different loop corrections on PT parameters in section~\ref{sec:PTparameters}, mentioning briefly the impact on GWs as well. We conclude in section~\ref{sec:conclusions}, where we also comment on the implications for the model with a real scalar. Technical details on sum-integrals, matching, running, the effective potential and the real scalar model are given in appendices~\ref{app:sumintegrals},  \ref{app:running}, \ref{app:matching}, \ref{app:effectivepotential} and \ref{app:othermodel}, respectively.

\section{Theoretical framework}
\label{sec:theory}
We consider a model consisting of a complex scalar field $\phi$ with global $U(1)_X$ charge $X=1$ and $N=3$ fermions $\psi_L$ and $\psi_R$ with charges $X=1$ and $X=0$, respectively, with the following Lagrangian in Minkowski space-time:
\begin{align}\label{eq:4DLagrangian}
    \mathcal{L}_{4} &= \partial_\mu\phi^\dagger \partial^\mu\phi - m^2 \phi^\dagger\phi - \lambda(\phi^\dagger\phi)^2 - \frac{c_{\phi^6}}{\Lambda^2}(\phi^\dagger\phi)^3
    \nonumber\\
    &\hphantom{=} + i\overline{\psi_L}\slashed{\partial}\psi_L + i\overline{\psi_R}\slashed{\partial}\psi_R - y (\phi \overline{\psi_L}\psi_R+\text{h.c.})\,,
\end{align}
where $\Lambda$ is some energy cut-off. We will work in TeV units throughout, and assume that $\Lambda=1$ TeV without loss of generality.

The high-temperature limit of this theory is described by a 3D EFT involving only the (loop-corrected) zeroth mode of $\phi$, which we call $\varphi$. The most general off-shell parametrisation of the corresponding Lagrangian up to dimension-6 operators (in 4D units), that we build using \texttt{ABC4EFT}~\cite{Li:2022tec}, reads as follows in Euclidean space:
\begin{align}\label{eq:3DLag}
	\mc{L}_{\text{EFT}} &= K_3\partial_{\mu} \varphi^{\dagger} \partial^{\mu} \varphi + m_3^2 \varphi^{\dagger} \varphi + \lambda_3 (\varphi^{\dagger} \varphi)^2 + c_{\varphi^6}(\varphi^{\dagger} \varphi)^3 + c^{(1)}_{\partial^2 \varphi^4} (\varphi^{\dagger} \varphi) (\partial_{\mu} \varphi^{\dagger} \partial^{\mu} \varphi)  \nn \\ 
	&\hphantom{=} + r^{(2)}_{\partial^2 \varphi^4} \left[(\varphi^{\dagger} \varphi) (\partial^2 \varphi^{\dagger} \varphi) + \text{h.c.}\right] + r^{(3)}_{\partial^2 \varphi^4} \left[i(\varphi^{\dagger} \varphi) (\partial^2 \varphi^{\dagger} \varphi) + \text{h.c.}\right] + r_{\partial^4 \varphi^2} \varphi^{\dagger} \partial^4 \varphi\,.
\end{align}

The WCs named with $r$ are redundant on-shell; that is, they can be removed via field redefinitions. Upon canonically normalising $\mathcal{L}_\text{EFT}$ (so that $K_3=1$), the equation of motion of $\varphi$ up to dimension 4 is
\begin{align}
	\partial^2 \varphi &= 
    m_3^2 \varphi + 2\lambda_3 (\varphi^{\dagger} \varphi)\varphi\,,
    %
\end{align}
from where the reduction of the redundant operators up to dimension 6 can be deduced:
\begin{align}
	\mc{R}^{(2)}_{\partial^2 \varphi^4} &= (\varphi^{\dagger} \varphi) (\partial^2 \varphi^{\dagger} \varphi) + \text{h.c.} = 
    2 m_3^2(\varphi^{\dagger} \varphi)^2 + 
    4\lambda_3
    (\varphi^{\dagger} \varphi)^3 \,, \\
	\mc{R}^{(3)}_{\partial^2 \varphi^4} &= i(\varphi^{\dagger} \varphi) (\partial^2 \varphi^{\dagger} \varphi) + \text{h.c.} = 0 \,, \\
	\mc{R}_{\partial^4 \varphi^2} &= (\varphi^{\dagger} \partial^4 \varphi) = 
    m_3^4 (\varphi^{\dagger} \varphi) + 
    4\lambda_3 m_3^2
    (\varphi^{\dagger} \varphi)^2 + 
    4\lambda^2_3
    (\varphi^{\dagger} \varphi)^3 \,.
\end{align}
Hence, the physical Lagrangian reads
\begin{align}\label{eq:3DLagphy}
	\mc{L}^{\text{phys}}_{\text{EFT}} = \partial_{\mu} \varphi^{\dagger} \partial^{\mu} \varphi + m'_3{}^2 \varphi^{\dagger} \varphi + \lambda'_3 (\varphi^{\dagger} \varphi)^2 + c'_{\varphi^6}(\varphi^{\dagger} \varphi)^3 + c'_{\partial^2 \varphi^4} (\varphi^{\dagger} \varphi) (\partial_{\mu} \varphi^{\dagger} \partial^{\mu} \varphi) \,,
\end{align}
and the above WCs are connected to those in Eq.~\eqref{eq:3DLag} by
%
%
\begin{align}
	\begin{aligned}
        &m'_3{}^2 = m_3^2 + m_3^4 r_{\partial^4 \varphi^2} \,, \quad \lambda'_3 = \lambda_3 + 2m_3^2 r^{(2)}_{\partial^2 \varphi^4} + 4\lambda_3 m_3^2 r_{\partial^4 \varphi^2} \,, \\
		&c'_{\varphi^6} = c_{\varphi^6} + 4\lambda_3 r^{(2)}_{\partial^2 \varphi^4} + 4\lambda^2_3 r_{\partial^4 \varphi^2} \,, \quad c'_{\partial^2 \varphi^4} = c_{\partial^2 \varphi^4} \,.
	\end{aligned}
\end{align}

\begin{figure}
    \centering
    \includegraphics[width=0.49\textwidth]{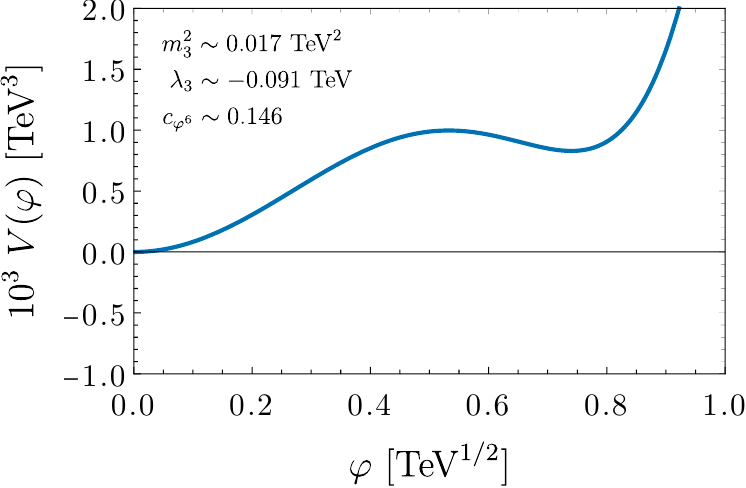}
    \includegraphics[width=0.49\textwidth]{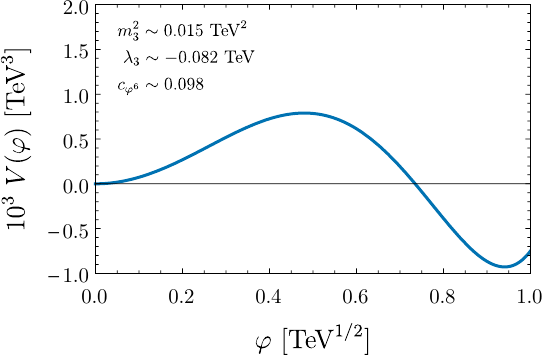}
    \caption{\it Leading scalar potential for different values of the 3D parameters. With a little abuse of notation, $\varphi$ here stands for the the real component of the complex scalar.}\label{fig:potential}
\end{figure}

This model is appealing in the following respects: (i) it very much resembles the Higgs sector of the SMEFT, while being significantly simpler; (ii) it is not as simple as the real scalar model~\cite{Chala:2024xll}, which presents no physical derivative interactions beyond the kinetic term; (iii) it can deliver two minima separated by a barrier while keeping a $\mathbb{Z}_2$ symmetry $\varphi\to-\varphi$, thus avoiding tadpole terms; see e.g. Fig.~\ref{fig:potential}.

We assume an $\mathcal{O}(1)$ Yukawa coupling.
For fixed $\lambda$, we characterise the parameter space of the model in terms of the physical squared mass ($m_P^2$) and vacuum expectation value ($v_P$) of $\phi$ at zero temperature:
\begin{align}
    m^2 = \frac{1}{4} (-m_P^2-2 \lambda v_P^2)\,,\quad
    \frac{c_{\phi^6}}{\Lambda^2} = \frac{1}{3 v_P^4} (m_P^2-2\lambda v_P^2)\,.
\end{align}
We take $\lambda$ as our power counting parameter. The rest of the couplings obey different power counting rules in different regions of the parameter space where there are PTs~\footnote{We check this trivially as follows. First, we determine the 3D parameters in the roughest approximation, which gives: $$m_3^2 \sim m^2+\frac{1}{4}y^2 T^2\,,\quad \lambda_3 \sim \lambda T\,,\quad c_{\varphi^6}\sim c_{\phi^6}T^2\,.$$ Then, we compute the values of $n_{m^2},n_y$ and $n_{c_{\phi^6}}$ that best fit the relations $(m^2/\pi)\sim (\lambda/\pi)^{n_{m^2}}$, $(y/\pi)\sim (\lambda/\pi)^{n_{y}}$ and $(c_{\phi^6}/\pi)\sim (\lambda/\pi)^{n_{c_{\phi^6}}}$.}. For SM-like values of $m_P^2$ and $v_P$, we have $(m / T)^2\sim y^2\sim\lambda$, $c_{\phi^6}\sim\lambda^2$~\cite{Croon:2020cgk,Camargo-Molina:2024sde,Chala:2025aiz}. However, for relatively large values of $\lambda$ and $m_P^2$, we have $(m/T)^2\sim y\sim\lambda$, $c_{\phi^6}\sim\lambda^2$. Within this latter parameter space region, all sum-integrals needed for computing the 3D EFT WCs to order $\lambda^3$, including 3-loop ones, are known; see Appendix \ref{app:sumintegrals}. Therefore, a fully consistent study of PTs at this order is achievable, which constitutes one further major advantage of this model.

Hence, in what follows, we consider two benchmark scenarios within this region of the parameter space:
\begin{align}
    \text{BP1}: (v_P, m_P^2) = (0.5\,\text{TeV}, 0.2\,\text{TeV}^2)\,,\quad \text{BP2}: (v_P, m_P^2) = (0.4\,\text{TeV}, 0.1\,\text{TeV}^2)\,,
\end{align}
and $y=0.9$ in both cases. (For relatively smaller values of $y$, there is no PT within this regime; for larger ones, SM-like power counting holds.) The region of the parameter space where a PT occurs is very tight (it occurs for $-0.5\lesssim\lambda\lesssim 0.3$), and shrinks for larger values of these parameters, which are moreover in tension with the assumed EFT cut-off $\Lambda=1$ TeV. 

The 4D and 3D parameters run following the corresponding renormalisation group equations (RGE), which depend in turn on the counterterms (CT). In the 4D theory, the latter are listed below:
\begin{align}\label{eq:4Dcounterterms}
    \delta K_\phi &= -\frac{3}{16\pi^2\epsilon}y^2-\frac{1}{128\pi^4\epsilon}\lambda^2,\\
    \delta m^2 &= \frac{1}{4\pi^2\epsilon}m^2\lambda + \frac{1}{64\pi^4}\left(-\frac{3}{\epsilon}m^2\lambda^2+\frac{7}{\epsilon^2}m^2\lambda^2\right)\,,\\
    \delta\lambda &= \frac{1}{8 \pi^2\epsilon}\left(5\lambda^2+\frac{9}{2}m^2\frac{c_{\phi^6}}{\Lambda^2}\right) + \frac{1}{64\pi^4}\left(-\frac{1}{\epsilon} 16\lambda^3+\frac{1}{\epsilon^2}25\lambda^3\right)\,, \label{eq:lambdacounterterm}\\
    \delta c_{\phi^6} &= \frac{3}{\pi^2\epsilon}\lambda\frac{c_{\phi^6}}{\Lambda^2}\,,\\
    \delta K_\psi &= -\frac{3}{32\pi^2\epsilon}y^2, \\
    \delta y &= 0 \,, \label{eq:ycounterterm}
\end{align}
where $K_\phi$ and $K_\psi$ stand for the kinetic terms of the scalar and the fermions, respectively. Here, $\delta K_\psi$ and $\delta y$ are computed only up to 1-loop because the 2-loop CTs of the fermionic interactions are irrelevant for the matching up to order $\lambda^3$. In the 3D theory, the CTs are as follows:
\begin{align}\label{eq:3Dcounterterms}
    \delta m_3^2 = \frac{1}{8\pi^2\epsilon} \lambda_3^2 \,,\quad \delta\lambda_3 = \frac{9}{8\pi^2\epsilon} \lambda_3 c_{\varphi^6}\,.
\end{align}
All others vanish to order $\lambda^3$.

We refer to Appendix~\ref{app:running} for the relevant diagrams and for the explicit computation of $\delta\lambda$ and $\delta\lambda_3$. Note that 1-loop integrals are not divergent in 3D, and that the squared mass does not renormalise at 3-loop either in 4D or in 3D. This is so because 3-loop diagrams necessarily scale with $\lambda c_{\phi^6}$ or with $\lambda^3$, which, contrary to $m^2$ ($m_3^2$), have energy dimensions $-2$ and $0$ ($1$ and $3$) in 4D (3D), respectively.

The perturbative solution to the 4D RGEs reads:
\begin{align}
    m^2(\mu) &= m^2 \left[1 + \frac{1}{8\pi^2}(4 \lambda + 3 y^2) \log \frac{\mu}{\Lambda} +\frac{1}{32\pi^4} \lambda^2 \left(14 \log^2 \frac{\mu}{\Lambda}  - 5 \log \frac{\mu}{\Lambda}\right)\right]\,,\label{eq:m2running}\\
    \lambda(\mu) &= \lambda \left[ 1 + \frac{1}{4 \pi^2} (5 \lambda + 3 y^2) \log \frac{\mu}{\Lambda} + \frac{5}{16 \pi^4} \lambda^2 \left( 5 \log^2 \frac{\mu}{\Lambda}  - 3 \log \frac{\mu}{\Lambda} \right) \right] \nonumber \\
    &+ \frac{9}{8 \pi^2} m^2 \frac{c_{\phi^6}}{\Lambda^2} \log \frac{\mu}{\Lambda} \,,\label{eq:lambdarunning}\\
    y(\mu) &= y \left[ 1 + \frac{3}{8 \pi^2} y^2 \log \frac{\mu}{\Lambda} + \frac{1}{256 \pi^4} \lambda^2 \log \frac{\mu}{\Lambda}\right]\,,
    \label{eq:yrunning}\\
    c_{\phi^6}(\mu) &= c_{\phi^6} \left[ 1 +  \frac{6}{ \pi^2} \lambda \log \frac{\mu}{\Lambda} \right] \,,\label{eq:c6running}
\end{align}
where the couplings on the right-hand side of the equations are implicitly evaluated at $\Lambda$.
We note that the running of the WCs above also encodes the running of $\phi$ and $\psi$, as they have been canonically normalised by their corresponding RGEs ---this is precisely why
$y$ runs despite $\delta y = 0$ in Eq.~\eqref{eq:ycounterterm}, before canonical normalisation. Some of these results can be cross-checked against \texttt{PyR@TE 3}~\cite{Sartore:2020gou}, with which we find full agreement.
    
In order to determine the EFT WCs in terms of the 4D couplings, we compute the hard region expansion of off-shell correlators involving the zeroth mode of $\phi$ in the Euclidean version of Eq.~\eqref{eq:4DLagrangian} in the static limit, $P^2=(0,\mathbf{p}^2)$, at order $\lambda^3$. This includes 1-loop diagrams for dimension-6 terms, up to 2-loop diagrams for the quartic and up to 3-loop diagrams for the squared mass. Subsequently, we match the result onto the tree-level counterpart in the EFT; see Appendix~\ref{app:matching}. This computation comprises the most demanding part of this work.

In order to simplify the expressions below, we introduce the following notation~\cite{Kajantie:1995dw}:
\begin{align}
    L_b &= L_b(\mu) \equiv 2 \log  \frac{e^{\gamma_E} \mu}{4 \pi T} \,,\quad L_f = L_f(\mu) \equiv 2 \log  \frac{e^{\gamma_E} \mu}{\pi T} \,,
\end{align}
where $\mu$ is the matching scale. All numerical constants and special functions that appear in the solution to sum-integrals are defined in Appendix \ref{app:sumintegrals}.

In the first place, we determine how the 4D zeroth mode of $\phi$ is related to $\varphi$ in the 3D EFT. This is given by the kinetic-term-matching equation:
\begin{equation}\label{eq:matchingK3}
    K_3 = 1 + \frac{3}{16 \pi^2} y^2 L_f + \frac{1}{768 \pi^4} \lambda^2 \left(19 + 12 L_b \right) \,.
\end{equation}
Then, we canonically normalise the 3D EFT through $\varphi \to \varphi / \sqrt{K_3}$. With a slight abuse of notation, we use the same names for the canonically normalised WCs and for the unnormalised WCs shown in Eq.~\eqref{eq:3DLag}. The rest of the (normalised) matching equations read:
\begin{align}
    \label{eq:matchingm32}
    m_3^2 &= m^2 + \lambda \left[\frac{1}{3}T^2 - \frac{1}{4 \pi^2} m^2 L_b + \frac{\zeta(3)}{32 \pi^4 T^2} m^4\right] + y^2 \left( \frac{1}{4}T^2 - \frac{3}{16 \pi^2} m^2 L_f \right) 
    \nonumber\\
    & + \frac{c_{\phi^6}}{\Lambda^2} \left( \frac{1}{8}T^4 - \frac{3}{16\pi^2} m^2 T^2 L_b\right)
    - \frac{1}{32 \pi^2} \lambda y^2 T^2 \left( 3 L_b + L_f \right) \nonumber \\
    & + \frac{1}{16 \pi^2} \lambda^2 \left[ T^2 \left( L_f - \frac{1}{3} L_b + 4 \log\pi - \frac{24 \zeta'(2)}{\pi^2} \blue{+ \frac{2}{\epsilon}} \right) + \frac{1}{4 \pi^2}m^2 \left( 7 L_b^2 + 5 L_b + \frac{89}{12} + \frac{4 \zeta(3)}{3} \right) \right] 
    \nonumber \\
    & + \frac{1}{16 \pi^2} \lambda\frac{c_{\phi^6}}{\Lambda^2} T^4 \left[\frac{3}{2} \left( L_b + L_f \right) +\frac{29}{10} - \frac{36 \zeta '(2)}{\pi ^2} + 360 \zeta'(-3) - 3 \gamma + 6 \log\pi + \blue{\frac{3}{\epsilon}} \right] \nonumber \\
    & + \frac{1}{128 \pi^4}\lambda^3 T^2 \left[ 2 C_{b} - 10 C_{s} - \frac{85}{3} L_b^2 - 5 L_f^2 + L_b \left( \frac{89}{3} + \frac{240 \zeta'(2)}{\pi^2} - \frac{80 \gamma}{3} \blue{- \frac{20}{\epsilon}} \right) \right. \nonumber\\
    & - L_f \left( \frac{29}{3} - \frac{80 \gamma}{3} + 40 \log\pi \right) - \frac{1}{9} \left( 313 \pi^2 + 509 \right) + \frac{4 \zeta(3)}{3} + \left( 41 - 20 \gamma \right) \frac{8 \zeta'(2)}{\pi^2} \nonumber \\
    & - 160 \zeta''(-1) + 8 \gamma \left( 19 \gamma - 2 \right) + \frac{992 \gamma_1}{3} + \frac{4}{3} \left( -29 + 80 \gamma - 60 \log\pi \right) \log\pi  \biggr]
    \,,
\end{align}
\begin{align}
    \label{eq:matchinglambda3}
    \lambda_3 &= \lambda T + \frac{c_{\phi^6}}{\Lambda^2} \left(\frac{3}{4}T^3 - \frac{9}{16\pi^2} m^2 T L_b\right) - \frac{5}{8 \pi^2}\lambda^2 \left[T L_b - \frac{\zeta(3)}{4 \pi^2 T} m^2 \right] - \frac{3}{8 \pi^2}\lambda y^2 T L_f\nonumber\\
    &+ \frac{9}{8 \pi^2} \lambda \frac{c_{\phi^6}}{\Lambda^2} T^3 \left[2 \log 2\pi - \frac{12\zeta'(2)}{\pi^2} \blue{+ \frac{1}{\epsilon}} \right] 
    + \frac{\lambda^3 T}{128 \pi^4}  \left[50 L_b^2 + 60 L_b + \frac{269}{3} + \frac{20 \zeta(3)}{3} \right] \,,
\end{align}
\begin{align}
    \label{eq:matchingc6}
    c_{\varphi^6} &= \frac{c_{\phi^6}}{\Lambda^2} T^2 - \frac{3}{\pi^2 }  \lambda \frac{c_{\phi^6}}{\Lambda^2} T^2 L_b + \frac{7 \zeta(3)}{24\pi^4} \lambda^3\,,\quad
    %
    c_{\partial^2\varphi^4}^{(1)} = r_{\partial^2\varphi^4}^{(2)} = -\frac{\zeta(3)}{48 \pi^4 T} \lambda^2\,;
    %
    %
\end{align}
while all others vanish at order $\lambda^3$.
\begin{figure}[t]
    \includegraphics[width=0.49\columnwidth]{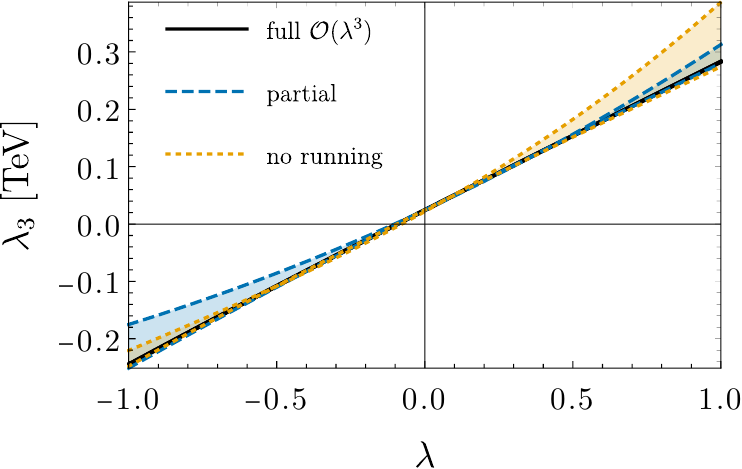}
    \includegraphics[width=0.49\columnwidth]{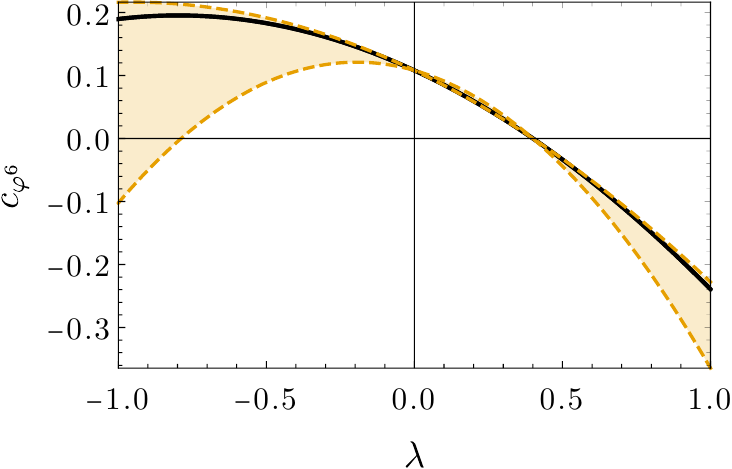}
    \caption{\it $\lambda_3$ (left) and $c_{\varphi^6}$ (right) for $T=\Lambda/\pi$ as a function of $\lambda$ in BP1, including both the running of 4D parameters and Coleman-Weinberg corrections (solid black), only the former (dashed blue) and none (dotted orange). The bands represent variations of the renormalization scale $\mu$ in the range $\mu\in [\overline{T}/2, 2\overline{T}]$, with $\overline{T} = \Lambda e^{-\gamma_E}$.}\label{fig:mudependence}
    \vspace{-1.5cm}
\end{figure}

Note that, upon replacing $\lambda_3$ and $c_{\varphi_6}$ in Eq.~\eqref{eq:3Dcounterterms} with their matching expressions in Eqs.~\eqref{eq:matchinglambda3} and \eqref{eq:matchingc6}, we obtain:
\begin{align}
    \delta m_3^2 &= \frac{1}{\epsilon} \left[ \frac{1}{16 \pi^2} \left( 2 \lambda^2 T^2 + 3 \lambda\frac{ c_{\phi^6}}{\Lambda^2} T^4 \right) - \frac{5}{32 \pi^4} \lambda^3 T^2 L_b \right] \,,\\
    \delta\lambda_3 &= \frac{9}{8 \pi^2 \epsilon} \lambda \frac{c_{\phi^6}}{\Lambda^2} T^3\,.
\end{align}
These are precisely the leftover divergences, shown in \textcolor{blue}{blue}, in Eqs.~\eqref{eq:matchingm32} and \eqref{eq:matchinglambda3}; all others are UV poles that are renormalised away. We remark that this is the result of a large number of cancellations, involving different loop orders, with and without CTs; see Appendix~\ref{app:matching}. It therefore constitutes an important cross-check for the matching. In particular, all double poles, of the form $1/\epsilon^2$, vanish.

The expressions above get further corrections from light loops, captured by the Coleman-Weinberg potential; see Appendix~\ref{app:effectivepotential}. Adding these to Eqs.~\eqref{eq:matchingm32} and \eqref{eq:matchinglambda3}, and taking into account the dependence of $m^2$, $\lambda$, $y$ and $c_{\phi^6}$ on $\mu$ given in Eqs.~\eqref{eq:m2running}--\eqref{eq:c6running}, the potential becomes renormalisation-scale invariant up to order $\lambda^3$; see Fig.~\ref{fig:mudependence}. (This is not exact in the case of $m_3^2$ because we neglect the 3-loop Coleman-Weinberg potential; however, the dependence of the renormalisation scale is tiny, and becomes generally imperceptible in numerical results). For $\lambda\sim -0.5$, ignoring both 4D and 3D running introduces renormalisation-scale dependence of about $20\,\%$ in $\lambda_3$ and of about $40\,\%$ in $c_{\varphi^6}$. The rest of the action is trivially independent of $\mu$.

As an example, let us show how $\lambda_3$, as determined from Eq.~\eqref{eq:matchinglambda3}, becomes scale-independent upon inserting the running of the UV WCs in Eqs.~\eqref{eq:m2running}--\eqref{eq:c6running} and the effective potential. Neglecting $\mathcal{O}(\lambda^4)$ corrections, we have:
\begin{align}
    \dot{\lambda}^{\rm full}_3  &\equiv \dot{\lambda}_3 + \dot{\lambda}_3^\text{eff} = \mu\frac{d}{d\mu}\left[ \lambda_3 - \frac{9}{4 \pi^2} \left(1 + 2\log\mu \right) c_{\varphi^6} \lambda_3 \right] \nonumber\\
    %
    %
    %
    %
    & = \dot{\lambda} T + \frac{3}{4} \frac{\dot{c}_{\phi^6}}{\Lambda^2} T^3 - \frac{9}{8 \pi^2} \frac{c_{\phi^6}}{\Lambda^2} m^2 T - \frac{5}{2 \pi^2} \dot{\lambda} \lambda T \left( \log\mu + \log \frac{e^\gamma}{4 \pi T} \right) - \frac{5}{4 \pi^2} \lambda^2 T \nonumber\\
    &\hphantom{=} - \frac{3}{4 \pi^2} \lambda y^2 T + \frac{1}{16 \pi^4} \lambda^3 T \left[50 \log\mu + \left(50 \log \frac{e^\gamma}{4 \pi T} + 15 \right) \right] - \frac{9}{2 \pi^2} \frac{c_{\phi^6}}{\Lambda^2} \lambda T^3 \nonumber\\
    &= \left( \frac{5}{4\pi^2}\lambda^2 + \frac{9}{8 \pi^2} \frac{c_{\phi^6}}{\Lambda^2} m^2 + \frac{3}{4 \pi^2} \lambda y^2 - \frac{15}{16 \pi^4} \lambda^3 \right) T + \frac{9}{2 \pi^2} \frac{c_{\phi^6}}{\Lambda^2} \lambda T^3 \nonumber\\
    &\hphantom{=}- \frac{9}{8 \pi^2} \frac{c_{\phi^6}}{\Lambda^2} m^2 T - \frac{5}{4 \pi^2} \lambda^2 T - \frac{25}{8 \pi^4} \lambda^3 T \left( \log\mu + \log \frac{e^\gamma}{4 \pi T} \right) \nonumber\\
    &\hphantom{=} - \frac{3}{4 \pi^2} \lambda y^2 T+ \frac{1}{16 \pi^4} \lambda^3 T \left[50 \log\mu + \left(50 \log \frac{e^\gamma}{4 \pi T} + 15 \right) \right] - \frac{9}{2 \pi^2} \frac{c_{\phi^6}}{\Lambda^2} \lambda T^3 \nonumber \\
    &= 0\,,
\end{align}
where $\lambda_3^{\text{eff}}$ is the effective potential contribution to the quartic coupling, which can be directly read from Eq.~\eqref{eq:effective_potential}, and the dot stands for $\mu \dfrac{d}{d\mu}$.

Non-local terms in the effective potential spoil the power counting in $\lambda$. In what follows, we include the 1-loop effective potential in the $\lambda^2$ and count $\log{\mu/m_3}$ as order $\lambda^0$ for the 2-loop part.  (These latter ones are in any case negligible; their effect is mainly to cancel the scale dependence of physical parameters.)

\section{Phase-transition parameters}
\label{sec:PTparameters}
The fundamental quantity to determine in any PT-related computation is the nucleation rate, which has the following form:
\begin{equation}
    \Gamma = A_\mathrm{stat} A_\mathrm{dyn} e^{-S_3[\varphi_c]}\,,
\end{equation}
where $S_3[\varphi_c]$ is the effective 3D action evaluated at the bounce solution \cite{Coleman:1977py}, of which it is an extremal, $A_\mathrm{stat}$ is the statistical pre-factor and $A_\mathrm{dyn}$ is the dynamical pre-factor~\cite{Ekstedt:2022tqk}. The first pre-factor accounts for equilibrium physics, and the latter captures non-equilibrium effects. In this work, we shall assume the high-temperature approximation $\Gamma \thickapprox T^4 e^{-S_3[\varphi_c]}$.

Since our aim is to quantify the effect of different matching corrections on PT parameters, we will for simplicity restrict to PTs in the real direction of $\varphi$. We take $\varphi = (\varphi_1  + i \varphi_2 )/\sqrt{2}$ and with a little abuse of notation, we use $\varphi$ to denote $\varphi_1$.

We compute $S_3[\varphi_c]$ using strict perturbation theory~\cite{Chala:2024xll}:
\begin{equation}
    S_3[\varphi_c] = S_3^{(0)}[\varphi_c^{(0)}] + S_3^{(1)}[\varphi_c^{(0)}]\,,
\end{equation}
where $S_3^{(0)}$ is the 3D action up to order $\lambda^2$ and $S_3^{(1)}$ stands for the $\mathcal{O}(\lambda^3)$ corrections. Likewise, $\varphi_c^{(0)}$ is the spherically-symmetric solution of the Euler-Lagrange equation~\cite{Coleman:1977py}
\begin{equation}
    \ddot{\varphi}_c^{(0)} + \frac{2}{r}\dot{\varphi}_c^{(0)} = V_3'(\varphi_c^{(0)})
\end{equation}
with boundary conditions $\dot{\varphi_c}^{(0)}(0)=0$ and $V^{(0)'}(\varphi_\infty^{(0)}) = 0$, where $\varphi_\infty^{(0)} \equiv \lim_{r \to \infty} \varphi^{(0)}(r)$.

We compute $\varphi_c^{(0)}$ using \texttt{FindBounce}~\cite{Guada:2020xnz}; see also Refs.~\cite{Wainwright:2011kj, Masoumi:2016wot, Athron:2019nbd, Sato:2019wpo,Hua:2025fap} for similar dedicated tools.
We assume that the PT takes place when the probability $\mathcal{P}\sim (M_{\rm Pl} / T)^4 \, e^{-S_3[\varphi_c]}$ for a single bubble to nucleate within a Hubble horizon volume is $\sim 1$. Numerically, this occurs when $S_3[\varphi_c]\sim 140$~\cite{Quiros:1999jp}. We denote by $T_*$ the temperature at which this holds.

Assuming that the Universe is radiation-dominated at the time of the PT, we define the following PT parameters, relevant for the production of GWs~\cite{Caprini:2019egz}.
\begin{itemize}[leftmargin=*, labelsep=0.5em]
 \item Strength parameter ($\alpha$). It is defined as the ratio of the trace anomaly difference of the energy momentum tensor between the symmetric and broken phases to the energy density of the radiation bath $\rho_r(T) = g(T) \pi^2 T^4/30$~\cite{Athron:2023xlk}: 
 \begin{equation}
  \alpha = \frac{1}{\rho_r(T_*)}\left.\Delta \left[ V_3(\varphi) - \frac{T}{4} \frac{d}{dT} V_3(\varphi) \right]\right|_{T_*}\,,
 \end{equation}
  with $g(T)$ being the number of relativistic degrees of freedom in the plasma at a given temperature. For the SM, at the time of the transition $g(T_*) = 106.75$ \cite{Weir:2017wfa}.
 \item Inverse duration ($\beta/H_*$). It is a characteristic timescale of the PT, corresponding to an exponentially growing transition rate as the temperature decreases (or equivalently, after linearising the bounce action with respect to the temperature) \cite{Caprini:2019egz}:
 \begin{equation}
  \frac{\beta}{H_*} = T_* \frac{d S_3[\varphi_c]}{d T}\bigg|_{T_*}\,.
 \end{equation}
 \item Terminal bubble wall velocity ($v_\omega$). In this work, we use the approximate formula \cite{Lewicki:2021pgr}
\begin{equation}
    v_w =
    \begin{cases}
        \sqrt{\frac{T \Delta V_3}{\alpha \rho_r}} & \text{for} \quad \sqrt{\frac{T \Delta V_3}{\alpha \rho_r}} < v_J(\alpha) \\
        1 & \text{for} \quad \sqrt{\frac{T \Delta V_3}{\alpha \rho_r}} \geq v_J(\alpha)
    \end{cases}\,,\quad  v_J = \frac{1}{\sqrt{3}} \frac{1 + \sqrt{3 \alpha^2 + 2 \alpha}}{1 + \alpha}\,,
    \label{eq:vw}
\end{equation}
where $\Delta V_3 = V_3(\varphi_T)$ is the difference in the potential between the phases.
\end{itemize}
The bubble wall velocity is determined from non-equilibrium processes, namely the interplay between the pressure between the scalar phases and the friction and back-reaction from the plasma. The precise computation of this parameter is a matter of ongoing study, as it is known to affect greatly the GW production from a FOPT; see e.g. \cite{Laurent:2022jrs, Ekstedt:2024fyq, Tian:2024ysd, Chen:2025ksr, Carena:2025flp}, and references therein.

\begin{figure}[t]
\includegraphics[width=0.48\textwidth]{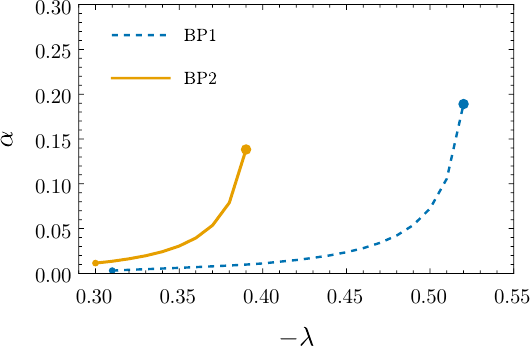}
\includegraphics[width=0.51\textwidth]{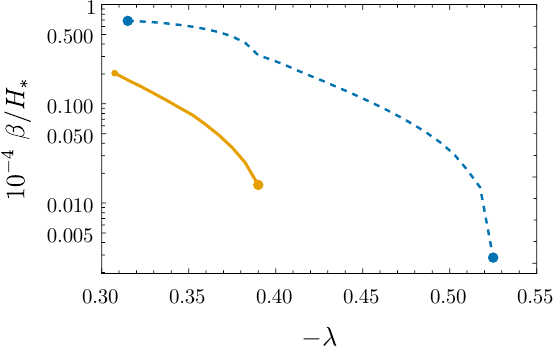}  
\caption{\it $\alpha$ (left) and $\beta/H_*$ (right) for BP1 (dashed blue) and BP2 (solid orange). The minimum and maximum values of $\lambda$ where there is a PT are marked.}\label{fig:leadingPTparameters}
\end{figure}

\begin{figure}[t]
    \includegraphics[width=0.49\columnwidth]{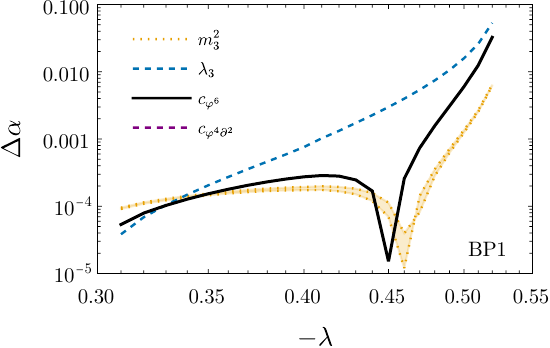}
    \includegraphics[width=0.49\columnwidth]{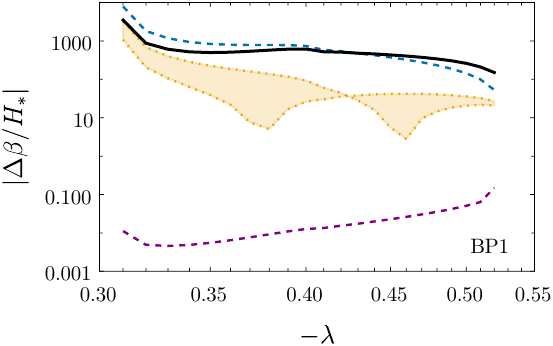}
    \includegraphics[width=0.49\columnwidth]{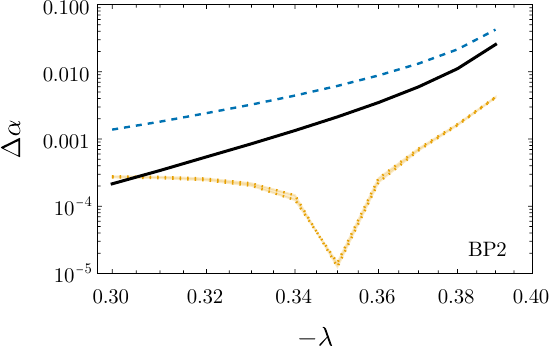}
    \hspace{0.1cm}
    \includegraphics[width=0.49\columnwidth]{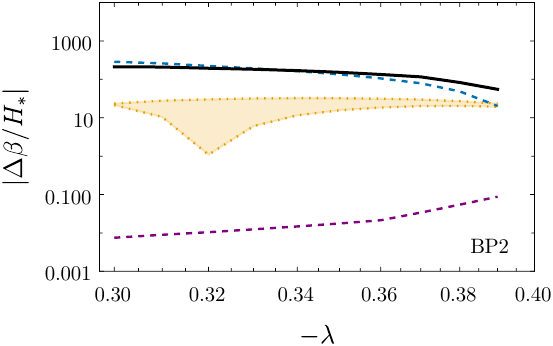}
    \caption{\it $\mathcal{O}(\lambda^3)$ contributions from all 3D physical operators to the strength parameter (left) and inverse duration (right) in BP1 (top) and BP2 (bottom). The bands represent variations of the renormalisation scale $\mu\in [\overline{T}/2, 2\overline{T}]$, with $\overline{T} = \Lambda e^{-\gamma_E}$.
    }
    \label{fig:PTparameters}
\end{figure}

To order $\lambda^2$, which in particular neglects effective interactions, we show $\alpha$ and $\beta/H_*$ in Fig.~\ref{fig:leadingPTparameters}. $T_*$ varies much less, ranging from $\sim 0.3 (0.2)$ to $0.15 (0.1)$ for BP1 (BP2).
Regarding $\mathcal{O}(\lambda^3)$ corrections, since our principal goal is to clarify the relative size of 1-loop effects of 3D effective operators versus 2-loop and 3-loop corrections to the mass and quartic coupling, we compare in Fig.~\ref{fig:PTparameters} only the $\mathcal{O}(\lambda^3)$ contributions to the above PT parameters~\footnote{We do so under the simplifying assumption that $T_*$ is not drastically modified by $\mathcal{O}(\lambda^3)$ corrections. This fails only when the overall correction to $\beta/H_*$ is so negatively large that no PT occurs, as we discuss further below.}

The spiky shape of the $m_3^2$ curves is due to the corresponding $\mathcal{O}(\lambda^3)$ corrections changing sign. From the plot, we infer that 1-loop corrections from $\varphi^6$ compete with the 2-loop quartic and far dominate over the 3-loop mass for sufficiently strong PTs (in particular. for those with $\alpha\gtrsim 0.1$, which are the ones that lead to observable GWs~\cite{Caprini:2019egz}).
Note that corrections to $\beta/H_*$ from $c_{\varphi^6}$ (and from $\lambda_3$) are negative. This can be understood as follows. In good approximation, $T_*$ and, therefore, the leading bounce are barely modified upon the introduction of $\mathcal{O}(\lambda^3)$ corrections. Consequently:
\begin{equation}
    \frac{\beta}{H_*} = T_* \frac{d}{d T}\left(S_3^{(0)}[\varphi_c^{(0)}]+S_3^{(1)}[\varphi_c^{(0)}]\right)\bigg|_{T_*} \thickapprox \frac{\beta^{(0)}}{H_*} + T_*\frac{dS_3^{(1)}[\varphi_c^{(0)}]}{dT}\bigg|_{T_*} \,,
\end{equation}
where $\beta^{(0)}/H_*$ is the value of $\beta/H_*$ computed without $\mathcal{O}(\lambda^3)$ corrections and the remainder is the correction we are interested in. Now, since $T_*$ and $\varphi_c^{(0)}$ are fixed, all the dependence on $T$ is encoded in the WCs. For the case of $c_{\varphi^6}$, we have:
\begin{align}
    \Delta\frac{\beta}{H_*} =  T_*\frac{dS_3^{(1)}[\varphi_c^{(0)}]}{dT}\bigg|_{T_*} \thickapprox 4\pi T_* \int dr r^2 [\varphi_c^{(0)}(r)]^6 ~\frac{1}{8} \frac{dc_{\varphi^6}}{d T}\bigg|_{T_*}\,,
\end{align}
which is negative for
\begin{align}\label{eq:negativebeta}
    \frac{dc_{\varphi^6}}{dT}\bigg|_{T_*} &= \frac{d}{dT}\left[\frac{6}{\pi^2} \left(\log{\frac{4\pi T}{\Lambda}}-\gamma_E\right) \lambda c_{\phi^6} T^2 + \frac{7 \zeta(3)}{24\pi^4}\lambda^3\right]\nonumber\\
    &= \frac{6}{\pi^2}\left(1-2\gamma_E+2\log{\frac{4\pi T}{\Lambda}}\right) \lambda c_{\phi^6}T <0 \Rightarrow T > \frac{e^{\gamma_E-\frac{1}{2}}}{4\pi}\Lambda < 0.1~\text{TeV}\,
\end{align}
for $\Lambda=1$ TeV, and therefore this correction is negative for all temperatures of interest.
This implies that, for $\lambda\gtrsim 0.5$, there is no PT within BP1, because the correction to $\beta/H_*$ makes it negative.

\begin{figure}[t]
    \includegraphics[width=.49\columnwidth]{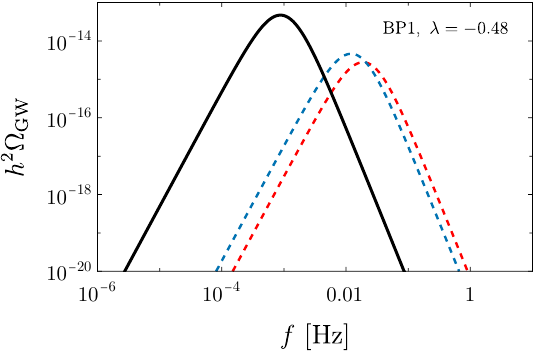}
    \includegraphics[width=.49\columnwidth]{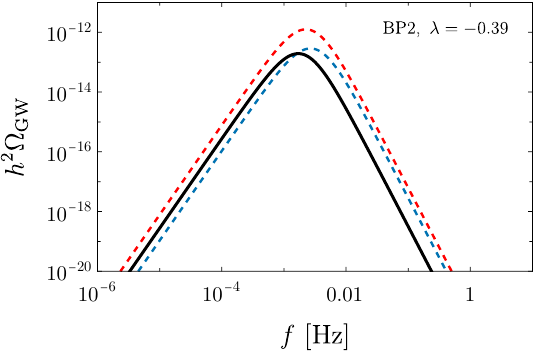}
    \caption{\it GW stochastic background generated during a PT computed at order $\lambda^2$ (dashed red), including 2-loop and 3-loop corrections from the mass and the quartic (dashed blue) and with 1-loop effective-operator corrections (solid black) in BP1 (left) and BP2 (right).}\label{fig:GWs}
\end{figure}

To conclude this analysis, we show the impact of $\mathcal{O}(\lambda^3)$ corrections in the GW spectrum of two different parameter space points in BP1 and BP2 computed using \texttt{PTPlot}~\cite{Caprini:2019egz, PTPlot}; see Fig.~\ref{fig:GWs}. It is apparent that 1-loop dimension-6 corrections can significantly dominate over 2-loop and 3-loop corrections on lower-dimensional interactions.

\section{Conclusions}
\label{sec:conclusions}
We have studied thermal-PT parameters within a model consisting of a complex scalar $\phi$ coupled to fermions, and in which the scalar potential exhibits two minima at zero temperature due to a $\phi^6$ interaction. We have done so within the framework of dimensional reduction, computing matching corrections to the mass, quartic and dimension-6 terms up to 3, 2 and 1 loops, respectively. This has been possible thanks to the quite unique characteristics of this model, that make that all 3-loop sum-integrals appearing in the process are known from hot QCD studies.

This way, we have been able to compare, for the first time, the relative importance of the different matching corrections, which, according to standard power counting, are in principle of the same order. We have found that, while 2-loop corrections to the quartic coupling compete with 1-loop corrections to $\varphi^6$, the latter generally dominate by a large margin over 3-loop corrections to the mass. 

\begin{figure}
    \includegraphics[width=0.32\columnwidth]{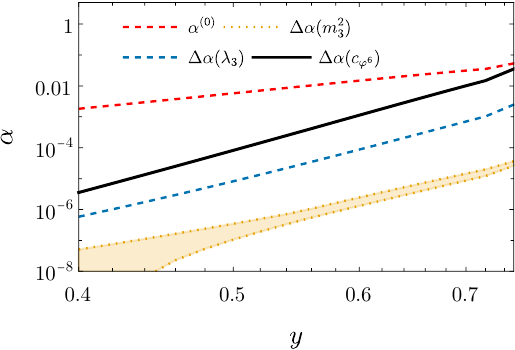}
    \includegraphics[width=0.32\columnwidth]{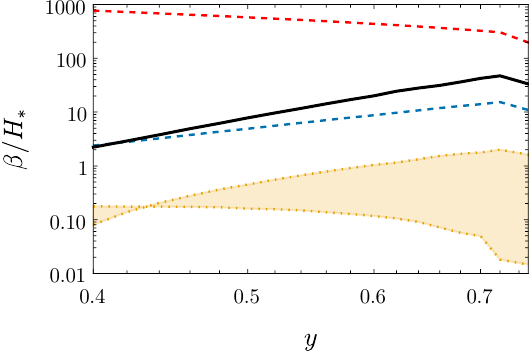}
    \includegraphics[width=0.32\columnwidth]{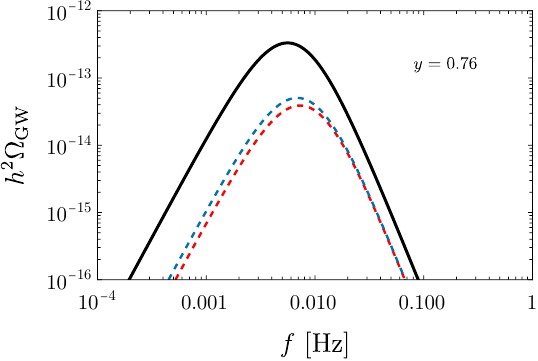}
    \caption{\it Strength parameter (left), inverse duration time (middle) and corresponding GW spectrum (right) in the model of Appendix~\ref{app:othermodel}, which extends Ref.~\cite{Chala:2024xll} with 2-loop matching corrections and running, for a benchmark point with $(m^2,\kappa,\lambda) = (0.02\,\mathrm{TeV}^2, -0.04\,\mathrm{TeV}, 0.1)$. Renormalisation-scale independent holds up to order $y^4$ ($y^6$) for the mass (quartic coupling and higher-dimensional operators). The bands represent variations of the renormalisation scale $\mu\in [\overline{T}/2, 2\overline{T}]$, with $\overline{T} = \Lambda e^{-\gamma_E}$. }\label{fig:realscalar}
\end{figure}

In order to further demonstrate the relevance of higher-order-operator corrections on PT parameters, we compute $\alpha$ and $\beta/H_*$, as well as the corresponding spectrum of GWs, within the model of Appendix~\ref{app:othermodel}, involving a real scalar singlet, a fermion and no dimension-6 terms. We include matching corrections up to 2-loops. (Unfortunately, 3-loop sum-integrals within this model are unknown.) The results are depicted in Fig.~\ref{fig:realscalar}. (Note that, unlike in Ref.~\cite{Chala:2024xll}, here we use $S_3[\varphi_c]\sim 140$, instead of $\sim 100$, as the nucleation criterion.) They show even more clearly the dominance of 1-loop corrections from dimension-6 terms. We find no reason to expect qualitatively different behaviour in other models of new physics.

Altogether, our results constitute the most robust evidence for the importance of dimension-6 operators compared to higher-loop corrections on lower-dimensional interactions in the 3D EFT for the study of strong PTs, particularly those detectable at current and future facilities. This does not necessarily imply that the high-temperature expansion is called into question, provided dimension-8-operator effects are sub-leading, which can be assessed following the methods of Ref.~\cite{Chala:2024xll}, and as we have ensured in all our results.

\section*{Acknowledgments}
We are indebted to York Schr\"oder for providing us with the analytic solution to two-loop fermionic sum-integrals. We are also grateful to Renato Fonseca for useful discussions. MC would like to thank the organisers and participants of the Portoroz 2025 workshop for valuable exchanges. We acknowledge support from the MCIN/AEI (10.13039/501100011033) and ERDF (grants
PID2021-128396NB-I00 and PID2022-139466NB-C22), from the Junta de Andaluc\'ia grants FQM 101 and P21-00199 and from Consejer\'ia de Universidad, Investigaci\'on
e Innovaci\'on, Gobierno de Espa\~na and Uni\'on Europea -- NextGenerationEU under grants AST22 6.5 and CNS2022-136024. This work has also been partially funded by MICIU/AEI/10.13039/501100011033 and ERDF/EU (grant PID2022-139466NB-C22). MC and LG are further supported by the RyC and FPU programs under contract numbers RYC2019-027155-I and FPU23/02026, respectively.

\appendix

\section{Sum-integrals}
\label{app:sumintegrals}

In what follows, we use a notation similar to that in Ref.~\cite{Laine_2020}, and we present all our results in the $\overline{\mathrm{MS}}$ scheme in dimensional regularisation, with $d = 3 - 2\epsilon$. We adopt the usual notation for sum-integrals:
\begin{equation}
    \sumintB{Q ~\mathrm{or}~\{Q\}} \equiv T \sum_{n=-\infty}^\infty \int_q\,\,,
\end{equation}
where $Q = (Q_0, \mathbf{q}) = (m_n, \mathbf{q})$ is a loop 4-momentum and $n$ labels the Matsubara modes running in the loop. The brackets denote a sum over fermionic modes ---for which we have $m_n = 2 \pi(n + \frac{1}{2}) T$---, while their absence means we sum over bosonic modes ---$m_n = 2 \pi n T$---.

Furthermore,
\begin{equation}
    \int_q \equiv \tilde{\mu}^{2\epsilon} \int \frac{d^{3-2\epsilon}q}{(2 \pi)^{3-2\epsilon}}\,,
    \label{eq:int not}
\end{equation}
where $\tilde{\mu}^2 \equiv e^{\gamma_E} \mu^2 / (4 \pi)$, $\mu$ being the $\overline{\mathrm{MS}}$ scale and $\gamma_E$ the Euler-Mascheroni constant.

\subsection{1-loop sum-integrals}
At 1-loop order, all bosonic sum-integrals, massive or massless, are known analytically. Since we expand sum-integrals in the scalar mass, we only need the massless cases, which read:
\begin{equation}
    \hat{I}_{\alpha}^{r} \equiv \sumintB{Q} \frac{Q_0^r}{Q^{2 \alpha}} = \tilde{\mu}^{2\epsilon} \frac{\left( 1 + (-1)^r \right) T}{(2 \pi T)^{2 \alpha - r - d}} \frac{\Gamma\left( \alpha - d/2 \right)}{(4 \pi)^{d/2} \Gamma\left( \alpha \right)} \zeta\left( 2 \alpha - r - d \right) \,,
\end{equation}
where $\Gamma(x)$ is the Euler gamma function and $\zeta(x)$ is the Riemann zeta function.

In the fermionic case, when the mass is non-zero, no analytic expressions are available. In the massless case, however, one can derive a simple relation with their bosonic counterpart. Scaling the spatial loop momentum $q \to 2 q$ and splitting the regularised infinite sum in odd and even integers, yields
\begin{equation}
    I_{\alpha}^{r} \equiv \sumintF{Q} \frac{Q_0^r}{Q^{2 \alpha}} = \left( 2^{2\alpha - r - d} - 1 \right) \hat{I}_{\alpha}^{r}\,.
\end{equation}

\subsection{2-loop sum-integrals}
All 2-loop sum-integrals in the matching can be written in terms of two bosonic or two fermionic loop momenta. The most general 2-loop bosonic sum-integral reads:
\begin{equation}
    \hat{I}_{\alpha \beta \gamma}^{r s} \equiv \sumintB{Q R} \frac{Q_0^r R_0^s}{Q^{2 \alpha} R^{2 \beta} (Q - R)^{2\gamma}}\,.
\end{equation}
To solve these, we use a recently developed algorithm~\cite{Davydychev:2023jto} that fully reduces any such structure to the 1-loop masters above.

Similarly, the most general 2-loop fermionic sum-integral reads:
\begin{equation}
    I_{\alpha \beta \gamma}^{r s} \equiv \sumintF{Q R} \frac{Q_0^r R_0^s}{Q^{2 \alpha} R^{2 \beta} ( Q - R )^{2\gamma}}\,.
\end{equation}
These are also known to factorize into 1-loop masters, however, in this case there exists no closed formula in the literature. Instead, these sum-integrals must be reduced on a case-by-case basis by means of symmetries induced by 4-momentum shifts and integration-by-parts  relations involving spatial momenta \cite{Nishimura:2012ee}.

By denoting $I_{\alpha \beta \gamma}^{00} \equiv I_{\alpha \beta \gamma}$ (resp. $\hat{I}_{\alpha \beta \gamma}^{00} \equiv \hat{I}_{\alpha \beta \gamma}$), we present below the specific 2-loop fermionic sum-integrals we need and their corresponding reductions:
\begin{align}
I_{111}^{\hphantom{00}} &= 0 \,, \\
I_{112}^{\hphantom{00}} &= \frac{1}{(d-2)(d-5)} \left( I_{220} - 2 I_{022} \right)\,, \\
I_{121}^{\hphantom{00}} &= I_{211} = -\frac{1}{(d-2)(d-5)} I_{220}\,, \\
I_{113}^{02} &= I_{113}^{20} = \frac{(d-3)(d-4)}{2 (d-2) (d-5) (d-7)} I_{022} + \frac{d-4}{d-7} I_{013}\,, \\
I_{131}^{02} &= I_{311}^{20} = \frac{d-4}{2 (d-2) (d-7)} \left( I_{220} + \frac{d-3}{d-5} I_{022} \right)\,, \\
%
%
I_{113}^{11} &= -\frac{d-4}{2 (d-2) (d-5) (d-7)} \left( I_{220} - 2 I_{022} \right) + \frac{d-4}{d-7} I_{013}\,, \\
%
%
I_{122}^{02} &= I_{212}^{20} = -\frac{d-4}{2 (d-2) (d-7)} \left( I_{220} + \frac{4}{d-5} I_{022} \right)\,, \\
I_{212}^{02} &= I_{122}^{20} = \frac{(d-4) (d^2 - 8d + 13)}{(d-2) (d-5) (d-7)} I_{022} + \frac{1}{d-7} \left( I_{031} - I_{130} \right)\,, \\
%
%
I_{122}^{11} &= I_{212}^{11} = \frac{d-4}{(d-2) (d-5) (d-7)} \left( I_{220} + \frac{d^2 - 8d + 11}{2} I_{022} \right)\,.
%
\end{align}

Finally, these can be straightforwardly reduced to 1-loop master integrals through:
\begin{align*}
    I_{\alpha \beta 0} &= I_{\alpha} I_{\beta}\,, \\
    I_{\alpha 0 \beta} &= I_{0 \alpha \beta} = I_{\alpha} \hat{I}_{\beta}\,,
\end{align*}
where in the second line we have used the shifts $R \to R - Q$ and $R \to -R$. Note that if $Q$ and $R$ are fermionic, shifting $R \to R - Q$ changes the nature of $R$ to bosonic.

\subsection{3-loop sum-integrals}
The evaluation of general 3-loop vacuum sum-integrals (bosonic, fermionic or mixed) is currently an open problem. For our present purpose, however, all cases have been conveniently solved in the context of hot QCD. 

The first subset that we find are trivial products of 1-loop masters:
\begin{equation}
    \sumintB{QRH} \frac{1}{Q^{2 \alpha} R^{2 \beta} H^{2 \gamma}} = \hat{I}_{\alpha} \hat{I}_{\beta} \hat{I}_{\gamma} \,.
\end{equation}
Others factorise into products of 1-loop masters and 2-loop sum-integrals, that we know how to further reduce to 1-loop masters. An example would be:
\begin{equation}
    \sumintB{QRH} \frac{1}{Q^{2 \alpha} R^{2 \beta} H^{2 \gamma} (R-H)^{2 \delta}} = \hat{I}_{\alpha} \hat{I}_{\beta \gamma \delta}\,.
\end{equation}
Finally, we also find non-trivial cases that are known analytically in the literature. In Eq.~(25) in Ref.~\cite{Moller_2010}, we find:
\begin{align}
     \sumintB{QRH} \frac{1}{Q^4 H^2 (Q-R)^2 (R-H)^2} &= \tilde{\mu}^{6 \epsilon} \left\{ \frac{T^2 (4\pi T^2)^{-3\epsilon}}{8(4\pi)^4\epsilon^2} \left[ 1+b_{21}\epsilon +b_{22}\epsilon^2 + \mathcal{O}(\epsilon^3) \right] \right\}\\
    b_{21} &= \frac{17}{6} + \gamma_E + 2 \frac{\zeta'(-1)}{\zeta(-1)} \nonumber \\
    b_{22} &=  \frac{131}{12} +\frac{31\pi^2}{36} +8\log 2\pi -\frac{9\gamma_E}{2} \nonumber \\
    &- \frac{15\gamma_E^2}{2} + (5+2\gamma_E) \frac{\zeta'(-1)}{\zeta(-1)} + 2 \frac{\zeta''(-1)}{\zeta(-1)} - 16 \gamma_1 \nonumber \\
    &+ \frac{4\zeta(3)}{9} + C_{b} \nonumber \,,
\end{align}
where $\gamma_1$ is one of the Stieltjes constants: $\zeta(1+\epsilon) = 1/\epsilon + \sum_{n=0}^\infty (-1)^n \gamma_n \epsilon^n /n!$; and the constant $C_{b} = -0.145652981107(4)$ is a sum of several dimensionless integrals that have been evaluated numerically. We have manually added the scale factor according to our definition of the integral measure.

Also, in Eq.~(2.36) in Ref.~\cite{Arnold_1994}, we find:
\begin{align}
    \sumintB{QRH} \frac{1}{Q^2 R^2 H^2 (Q+R+H)^2} = \frac{1}{(4 \pi)^2} \left( \frac{T^2}{12} \right)^2 \bigg[\frac{6}{\epsilon}& + 36 \log\frac{\mu}{4 \pi T} - 12 \frac{\zeta'(-3)}{\zeta(-3)}  \nonumber \\
    & +  48 \frac{\zeta'(-1)}{\zeta(-1)} + \frac{182}{5} \bigg] + \mathcal{O}(\epsilon)\,,
\end{align}
which assumes the same integral measure we use. Note that the result is expressed in terms of the $\overline{\mathrm{MS}}$ scale $\mu$, and not in terms of $\tilde{\mu}$.

Finally, from Eq.~(2.15) in Ref.~\cite{Schroder_2012} we read:
\begin{align}
    &\sumintB{QRH} \frac{1}{Q^2 R^2 H^2 (Q-R)^2 (Q-H)^2} \nonumber \\
    &= \tilde{\mu}^{6 \epsilon} \left\{-\frac{1}{4} \frac{T^2}{(4 \pi)^4} \frac{(4 \pi e^{\gamma_E} T^2)^{-3 \epsilon}}{\epsilon^2} \left[ 1 + v_1 \epsilon + v_2 \epsilon^2 + \mathcal{O}(\epsilon^3)\right] \right\}\,; \\
    &v_1 = \frac{4}{3} + 4 \gamma_E + 2 \frac{\zeta'(-1)}{\zeta(-1)} \,, \nonumber\\
    &v_2 = \frac{1}{3} \left[ 46 - 16 \gamma_E^2 + \frac{45 \pi^2}{4} + 24 \log^2 2 \pi - 104 \gamma_1 - 8 \gamma_E \right. \nonumber \\
    &\hphantom{v_2 = } \left. - 24 \gamma_E \log 2 \pi + 16 \gamma_E \frac{\zeta'(-1)}{\zeta(-1)} + 24 \frac{\zeta'(-1)}{\zeta(-1)} + 2 \frac{\zeta''(-1)}{\zeta(-1)} \right] + C_{s} \,, \nonumber
\end{align}
where the constant $C_{s} = - 38.5309$ is a sum of several dimensionless integrals that have been evaluated numerically.

\section{Running of $\lambda$ and $\lambda_3$}
\label{app:running}
We start describing the CT Lagrangian in the 4D theory:
\begin{align}\label{eq:4DLagrangian CT}
    \mathcal{L}_{4, \rm ct} &= \delta K_\phi \partial_\mu\phi^\dagger \partial^\mu\phi - \delta m^2 \phi^\dagger\phi - \delta\lambda(\phi^\dagger\phi)^2 - \frac{\delta c_{\phi^6}}{\Lambda^2}(\phi^\dagger\phi)^3\,,
    \nonumber\\
    &\hphantom{=} + i \delta K_\psi \left( \overline{\psi_L}\slashed{\partial}\psi_L + i \overline{\psi_R}\slashed{\partial}\psi_R \right) - \delta y (\phi \overline{\psi_L}\psi_R+\text{h.c.})\,,
\end{align}
as well as the counterpart in the 3D EFT:
\begin{align}\label{eq:3DLag CT}
	\mc{L}_{\text{EFT}, \rm ct} &= \delta Z_\varphi (\partial_{\mu} \varphi)^{\dagger} (\partial^{\mu} \varphi) + \delta m_3^2 \varphi^{\dagger} \varphi + \delta\lambda_3 (\varphi^{\dagger} \varphi)^2 + \delta c_{\varphi^6}(\varphi^{\dagger} \varphi)^3 \nn\\
    &\hphantom{=} + \delta c^{(1)}_{\partial^2 \varphi^4} (\varphi^{\dagger} \varphi) (\partial_{\mu} \varphi^{\dagger} \partial^{\mu} \varphi) + \delta r^{(2)}_{\partial^2 \varphi^4} \left[(\varphi^{\dagger} \varphi) (\partial^2 \varphi^{\dagger} \varphi) + \text{h.c.}\right] \nn\\
    &\hphantom{=} + \delta r^{(3)}_{\partial^2 \varphi^4} \left[i(\varphi^{\dagger} \varphi) (\partial^2 \varphi^{\dagger} \varphi) + \text{h.c.}\right] + \delta r_{\partial^4 \varphi^2} (\varphi^{\dagger} \partial^4 \varphi)\,.
\end{align}

\begin{figure}[t]
    \centering
    \includegraphics[width=0.19\textwidth]{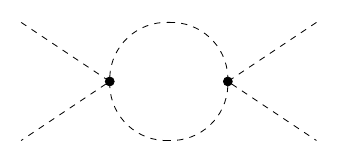} 
    \includegraphics[width=0.19\textwidth]{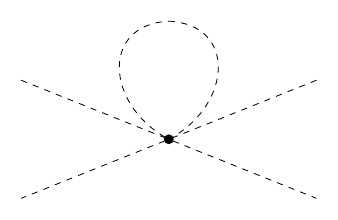}
    \includegraphics[width=0.19\textwidth]{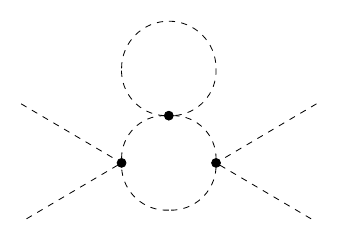}
    \includegraphics[width=0.19\textwidth]{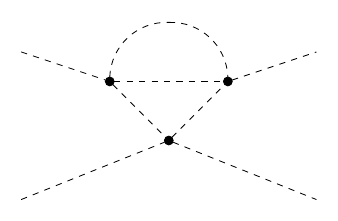}
    \includegraphics[width=0.19\textwidth]{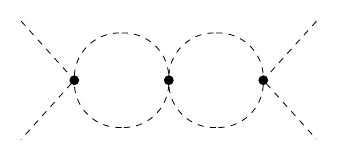}
    \includegraphics[width=0.19\textwidth]{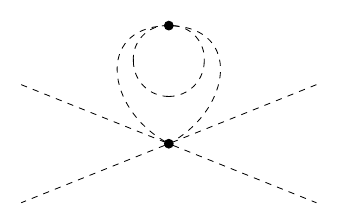}
    \includegraphics[width=0.19\textwidth]{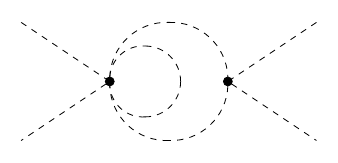}
    \includegraphics[width=0.19\textwidth]{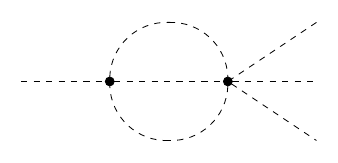}
    \includegraphics[width=0.19\textwidth]{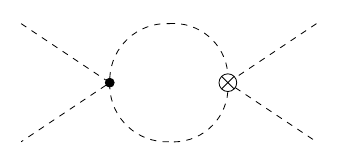}
    \caption{\it Relevant 1-loop and 2-loop diagrams for the running of the 4-point functions in the 4D theory and the 3D EFT. The cross with one circle denotes the 1-loop CT.}\label{fig:running4}
\end{figure}

As a clarifying example, let us explicitly derive first $\delta \lambda$ and then $\delta \lambda_3$ to order $\lambda^3$. The relevant diagrams are shown in Fig.~\ref{fig:running4}, among which only the sunset diagram, that is, the 8th diagram in Fig.~\ref{fig:running4}, contributes to $\delta \lambda_3$. We work in the $\rm \overline{MS}$-scheme in dimensional regularisation.

In the 4D theory, we split the 4-point function with the scalar in the external legs in loop orders as
\begin{equation}
    \Gamma_{\phi\phi\phi\phi} = \Gamma_{\phi\phi\phi\phi}^{(0)} + \Gamma_{\phi\phi\phi\phi}^{(1)} + \Gamma_{\phi\phi\phi\phi}^{(2)} \,,
\end{equation}
and we present each piece separately, after simplifying the traces of gamma matrices and removing higher order terms. We do the same for the CT correlators, which we denote $\Gamma_{\phi\phi\phi\phi, \rm ct}^{(\ell)}$. For convenience, let us also define a pole-subtracting operator $\mathcal{K}$ with the property that
\begin{equation}
    \mathcal{K}\left( a_0 + \sum_{k=1}^n \frac{a_k}{\epsilon^k} \right) = \sum_{k=1}^n \frac{a_k}{\epsilon^k}\,,
\end{equation}
with $a_k \in \mathbb{C}, k=0, 1, 2, \dots$ being $\epsilon$-independent.

Since the tree-level part is not divergent, let us start with the 1-loop, that reads:
\begin{equation}
    \Gamma_{\phi\phi\phi\phi}^{(1)} = 40 \lambda^2 \int_q \frac{1}{(q^2 - m^2)^2} + 36 \frac{c_{\phi^6}}{\Lambda^2} \int_q \frac{1}{q^2 - m^2} \,,
\end{equation}
where we use the same notation as in Eq. \eqref{eq:int not} but in $d=4-2\epsilon$.

The evaluation of these 1-loop integrals in dimensional regularization is straightforward, and it can be found in any standard QFT textbook (see e.g. Appendix B in Ref. \cite{Schwartz:2014sze}). The result is:
\begin{equation}
    \mathcal{K} \left(\Gamma_{\phi\phi\phi\phi}^{(1)}\right) = \frac{i}{4 \pi^2 \epsilon} \left( 10 \lambda^2 + 9 m^2 \frac{c_{\phi^6}}{\Lambda^2} \right)\,.
\end{equation}

The corresponding CT diagram is
\begin{equation}
    \Gamma_{\phi\phi\phi\phi, \rm ct}^{(1)} = -4 i \delta \lambda^{(1)}\,,
\end{equation}
so, by definition, 
\begin{equation}
    \delta\lambda^{(1)} = \frac{1}{8 \pi^2} \left( 5 \lambda^2 + \frac{9}{2} m^2 \frac{c_{\phi^6}}{\Lambda^2} \right)\,.
\end{equation}

At 2-loop level, we have:
\begin{align}
    \Gamma_{\phi\phi\phi\phi}^{(2)} =& 288 i \lambda \frac{c_{\phi^6}}{\Lambda^2} \int_{q, k} \frac{1}{(q^2-m^2)(k^2-m^2)[(q+k)^2-m^2]} \nonumber \\
    &+ 144 i \lambda \frac{c_{\phi^6}}{\Lambda^2} \int_{q, k} \frac{1}{(q^2-m^2)(k^2-m^2)^2} 
    + 720 i \lambda \frac{c_{\phi^6}}{\Lambda^2} \int_{q, k} \frac{1}{(q^2-m^2)^2 (k^2-m^2)} \nonumber \\
    &+ 512 i \lambda^3 \int_{q, k} \frac{1}{(q^2-m^2)(k^2-m^2)[(q+k)^2-m^2]^2} \nonumber \\
    &+ 144 i \lambda^3 \int_{q, k} \frac{1}{(q^2-m^2)^2 (k^2-m^2)^2}
    + 320 i \lambda^3 \int_{q, k} \frac{1}{(q^2-m^2)^3 (k^2-m^2)}\,.
\end{align}
Using the known formulae for 2-loop tadpole integrals in Ref. \cite{Davydychev:1992mt}, we find that the divergent part is, neglecting higher orders:
\begin{equation}
    \mathcal{K}\left(\Gamma_{\phi\phi\phi\phi}^{(2)}\right) = \frac{i \lambda^3}{16 \pi^4} \left[ \left( - 6 - 100 \log \frac{\mu}{m} \right) \frac{1}{\epsilon} - \frac{25}{\epsilon^2} \right]\,.
\end{equation}

Now, for the sake of simplicity, let us assume that we already know the rest of the 1-loop CTs, that can be easily obtained as we just did for $\delta\lambda^{(1)}$. Though they do not appear explicitly in this example, let us however note that loops with massless fermions must be computed with care, as they can cause the mixing of UV and IR divergences. In order to avoid this, we introduce a spurious mass $m_f$ that we take to zero once all integrals have been evaluated. 

From the CT diagrams, the only terms to $\mathcal{O}(\lambda^3)$ are:
\begin{equation} \label{eq:pole1}
    \Gamma_{\phi\phi\phi\phi, \rm ct}^{(2)} = 80 \delta\lambda^{(1)} \lambda \int_q \frac{1}{(q^2-m^2)^2} - 4 i \delta \lambda^{(2)}\,,
\end{equation}
which yields, 
\begin{equation} \label{eq:pole2}
    \mathcal{K} \left(\Gamma_{\phi\phi\phi\phi, \rm ct}^{(2)}\right) = \frac{i \lambda^3}{16 \pi^4} \left[ \left(-10 +  100 \log  \frac{\mu}{m}  \right) \frac{1}{\epsilon} + \frac{50}{\epsilon^2} \right] - 4 i \delta \lambda^{(2)} \,.
\end{equation}
Summing Eqs. \eqref{eq:pole1} and \eqref{eq:pole2}, we find
\begin{equation}
    \delta\lambda^{(2)} = \frac{\lambda^3}{64 \pi^4} \left( - \frac{16}{\epsilon} + \frac{25}{\epsilon^2} \right)\,.
\end{equation}
Thus, we recover the CT shown in Eq.~\eqref{eq:lambdacounterterm} in the main text.

For the computation of $\delta \lambda_3$, the process is analogous, but simpler. We split the 4-point function in loop orders as
\begin{equation}
    \Gamma_{\varphi\varphi\varphi\varphi} = \Gamma_{\varphi\varphi\varphi\varphi}^{(0)} + \Gamma_{\varphi\varphi\varphi\varphi}^{(1)} + \Gamma_{\varphi\varphi\varphi\varphi}^{(2)} \,.
\end{equation}
and we shall present all results in Euclidean space.

We know that $\Gamma_{\varphi\varphi\varphi\varphi}^{(1)}$ cannot be divergent, as there are no divergent tadpole integrals at 1-loop order in $d=3$. Therefore, 
\begin{equation}
    \delta \lambda_3^{(1)} = 0\,.
\end{equation}

At 2-loop order, we also know that the only divergent tadpole integral is the one associated to the sunset diagram. Therefore, focusing on this type only, we have:
\begin{equation}
    \Gamma_{\varphi\varphi\varphi\varphi}^{(2)} = - 288 \lambda_3 c_{\varphi^6} \int_{q, r} \frac{1}{(q^2 + m_3^2)(r^2 + m_3^2)[(q+r)^2 + m_3^2]} + \dots
\end{equation}
where the ellipses include all other non-divergent contributions. Again, using the known formulae for massive 2-loop integrals, we obtain:
\begin{equation}
    \mathcal{K} \left(\Gamma_{\varphi\varphi\varphi\varphi}^{(2)} \right) = \frac{9}{2 \pi^2} \lambda_3 c_{\varphi^6}\,.
\end{equation}

Since there are no 1-loop CTs, we must only compute the tree-level insertion of 2-loop CTs, which yields:
\begin{equation}
    \Gamma_{\varphi\varphi\varphi\varphi, \rm ct}^{(2)} = - 4 \delta \lambda_3^{(2)} + \dots
\end{equation}

We therefore obtain:
\begin{equation}
    \delta \lambda_3^{(2)} = \frac{9}{8 \pi^2} \lambda_3 c_{\varphi^6}\,,
\end{equation}
as shown in Eq. \eqref{eq:3Dcounterterms} in the main text.

\section{Matching}
\label{app:matching}
As stated in the main text, we have the following power counting in the 4D theory:
\begin{equation}
     y \sim \frac{m^2}{T^2} \sim \frac{|\mathbf{p}|^2}{T^2} \sim \lambda\,,\quad \frac{c_{\phi^6}}{\Lambda^2} \sim \lambda^2\,.
     \label{eq:power counting 1}
\end{equation}
In order to perform the hard region expansion of 4D correlators, we expand in powers of $|\mathbf{p}|^2/T^2$ and $m^2/T^2$ by iterating the following identity:
\begin{equation}
    \frac{1}{(Q+P)^2 + m^2} = \frac{1}{Q^2 + m^2} \left[ 1 - \frac{P^2 + 2 (Q \cdot P) + m^2}{(Q+P)^2 + m^2} \right] \,,
\end{equation}
and Taylor-expanding in $m^2/Q^2$ up to the needed order in $\lambda$.

We use the following tensor reduction formulae to simplify different tensor structures to scalar integrals:
\begin{align} \label{eq:tensor reds}
    q_i r_j &= \frac{\mathbf{q} \cdot \mathbf{r}}{d} \delta^{ij} \,, \nonumber\\
    q_i r_j r_k r_l &= \frac{|\mathbf{r}|^2  \left(\mathbf{q} \cdot \mathbf{r} \right)}{d^2 + 2 d} \left( \delta_{ij} \delta_{kl} + \delta_{ik} \delta_{jl} + \delta_{il} \delta_{jk} \right) \,, \nonumber\\
    q_i q_j r_k r_l &= \frac{\left( \delta_{ik} \delta_{jl} + \delta_{il} \delta_{jk}\right) \left[ d \left( \mathbf{q} \cdot \mathbf{r} \right)^2 - |\mathbf{q}|^2 |\mathbf{r}|^2\right]+ \delta_{ij} \delta_{kl} \left[(d+1) |\mathbf{q}|^2 |\mathbf{r}|^2 - 2 \left( \mathbf{q} \cdot \mathbf{r} \right)^2 \right]}{d (d-1) (d+2)}\,,
\end{align}
where $q_i$ and $r_i$ are spatial 3-momenta. The same reductions in the case of one single independent loop momentum can be read from the ones above by simply setting $q_i = r_i$.

Finally, we apply specific linear shifts of loop momenta and algebraic identities to rewrite all scalar vacuum sum-integrals as master sum-integrals that are left to evaluate; see Appendix \ref{app:sumintegrals}. 
\begin{figure}[t]
    \centering
    \includegraphics[width=0.19\textwidth]{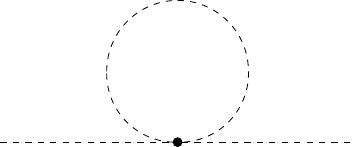}
    \includegraphics[width=0.19\textwidth]{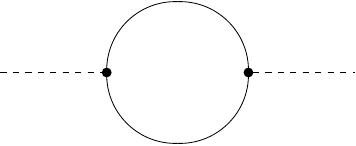}
    \includegraphics[width=0.19\textwidth]{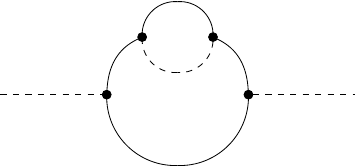}
    \includegraphics[width=0.19\textwidth]{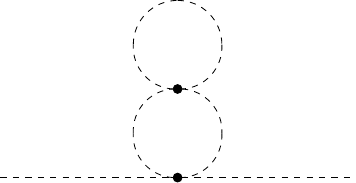}
    \includegraphics[width=0.19\textwidth]{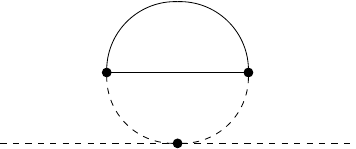}
    \includegraphics[width=0.19\textwidth]{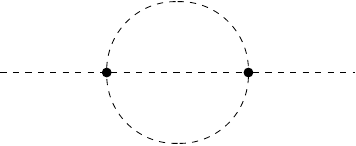}
    \includegraphics[width=0.19\textwidth]{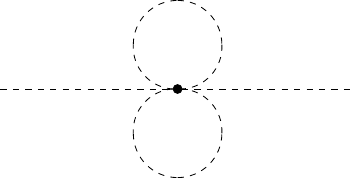}
    \includegraphics[width=0.19\textwidth]{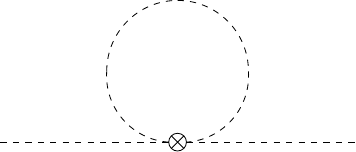}
    \includegraphics[width=0.19\textwidth]{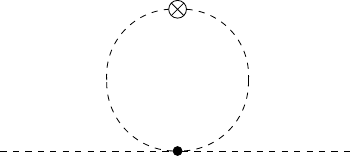}
    \caption{\it 1-loop and 2-loop diagrams for the 2-point function. The crosses with one circle denote the 1-loop CTs.}\label{fig:2-loops}
    \vspace{0.7cm}
\end{figure}
\begin{figure}[t]
    \includegraphics[width=0.19\textwidth]{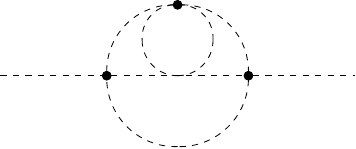}
    \includegraphics[width=0.19\textwidth]{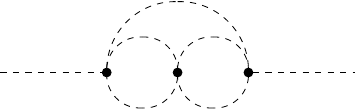}
    \includegraphics[width=0.19\textwidth]{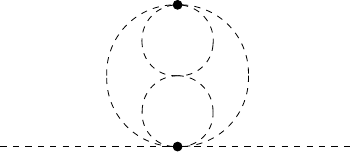}
    \includegraphics[width=0.19\textwidth]{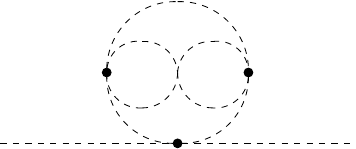}
    \includegraphics[width=0.19\textwidth]{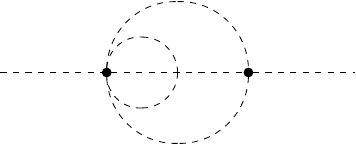}
    \includegraphics[width=0.19\textwidth]{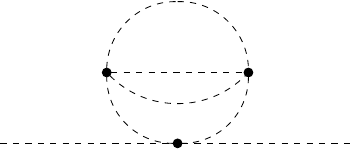}
    \includegraphics[width=0.19\textwidth]{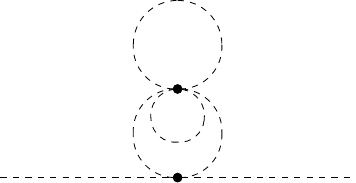}
    \includegraphics[width=0.19\textwidth]{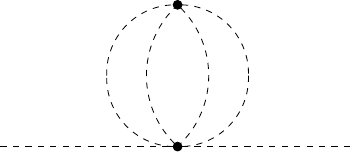}
    \includegraphics[width=0.19\textwidth]{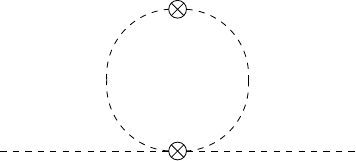}
    \includegraphics[width=0.19\textwidth]{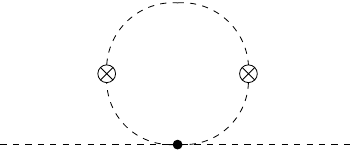}
    \includegraphics[width=0.19\textwidth]{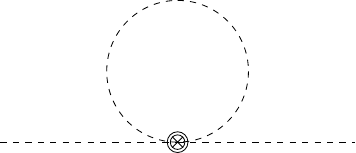}
    \includegraphics[width=0.19\textwidth]{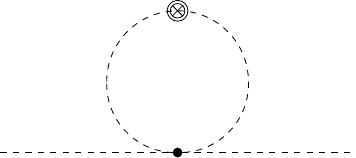}
    \includegraphics[width=0.19\textwidth]{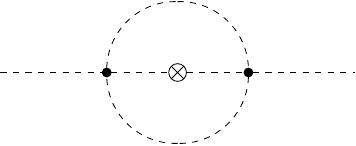}
    \includegraphics[width=0.19\textwidth]{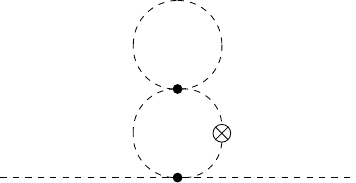}
    \includegraphics[width=0.19\textwidth]{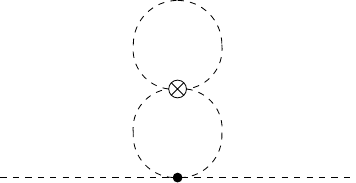}
    \includegraphics[width=0.19\textwidth]{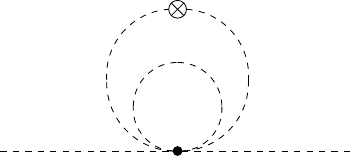}
    \includegraphics[width=0.19\textwidth]{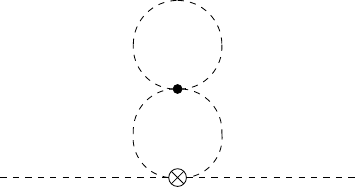}
    \includegraphics[width=0.19\textwidth]{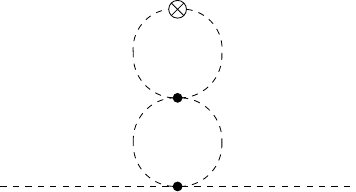}
    \includegraphics[width=0.19\textwidth]{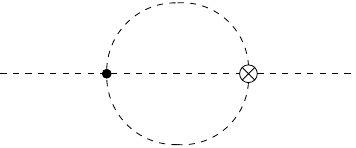}
    \includegraphics[width=0.19\textwidth]{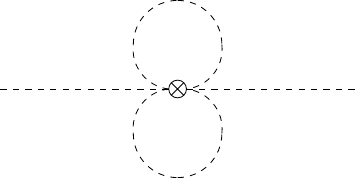}
    \caption{\it 3-loop diagrams for the 2-point function. The crosses with one circle denote the 1-loop CTs and the crosses with two circles denote the 2-loop CTs.}
    \label{fig:3-loops}
\end{figure}
As an explicit example of the procedure above, we show here how to obtain the matching equation for the kinetic term of the scalar in the EFT up to order $\lambda^3$, that is, the result in Eq.~\eqref{eq:matchingK3}. The relevant diagrams for the 2-point function are shown in Figs.~\ref{fig:2-loops} and \ref{fig:3-loops}.

Let us first note that, while the 3-loop diagrams represented in Fig.~\ref{fig:3-loops} contribute to the matching of the effective mass, they do not contribute to the kinetic term up to order $\lambda^3$. This is because the vertex prefactor of all 3-loop diagrams already add up to this order, so their expansion in external momentum up to $|\mathbf{p}|^2$ is at least $\mathcal{O}(\lambda^4)$ according to our power counting.

We split the 2-point function with the scalar zero mode in the external legs in loop orders as
\begin{equation}
    \Gamma_{\phi_0 \phi_0} \equiv \Gamma_{\phi_0 \phi_0}^{(0)} + \Gamma_{\phi_0 \phi_0}^{(1)} + \Gamma_{\phi_0 \phi_0}^{(2)}\,,
\end{equation}
and we present each piece separately in Euclidean space and after simplifying the traces of gamma matrices in fermionic sum-integrals.

The tree-level trivially reads:
\begin{equation}
    \Gamma_{\phi_0 \phi_0}^{(0)} = -P^2 - m^2 \,.
\end{equation}

The 1-loop piece, which corresponds to the first two diagrams in Fig. \ref{fig:2-loops} together with tree-level insertions of 1-loop CTs, is:
\begin{equation}
    \Gamma_{\phi_0 \phi_0}^{(1)} = - 4 \lambda \sumintB{Q} \frac{1}{Q^2 + m^2} + 6 y^2 \sumintF{Q} \frac{Q^2 - (P \cdot Q)}{Q^2 (Q-P)^2} - \delta K_\phi^{(1)} P^2 - \delta m^2{}^{(1)}\,.
\end{equation}
Since we only want to determine the kinetic term, we shall only focus on the pieces which depend on external momentum. Applying the hard region expansion up to $\mathcal{O}(\lambda^3)$, we get terms proportional to $P^2$, that contribute to the kinetic term, and some momentum-independent terms. Removing odd terms in $Q$, that are vanishing, and collecting all momentum-independent terms in the ellipses, we obtain:
\begin{equation}
    \Gamma_{\phi_0 \phi_0}^{(1)} = 6 y^2 \sumintF{Q} \left[ 2\frac{(P \cdot Q)^2}{Q^6} - \frac{P^2}{Q^4} \right] - \delta K_\phi^{(1)} P^2 + \dots
\end{equation}
Now, since $P = (0, \mathbf{p})$, $P \cdot Q = \mathbf{p} \cdot \mathbf{q}$ and we can use the tensor reduction formulae in Eqs.~\eqref{eq:tensor reds} to remove the mixed scalar product. Then, rewriting $|\mathbf{q}|^2 = Q^2 - Q_0^2$, we finally express the result in terms of 1-loop vacuum sum-integrals:
\begin{equation}
    \Gamma_{\phi_0 \phi_0}^{(1)} = 6 y^2 |\mathbf{p}|^2 \left[ \frac{2}{d} \left(I_2^0 - I_3^2 \right) - I_2^0 \right] - \delta K_\phi^{(1)} |\mathbf{p}|^2  + \dots
\end{equation}

Moving on to the 2-loop piece, keeping terms up to $\mathcal{O}(\lambda^3)$, and again including all momentum-independent terms in the ellipses, we only have:
\begin{align}
    \Gamma_{\phi_0 \phi_0}^{(2)} = 8 \lambda^2 \sumintB{Q R} \frac{1}{(Q^2 + m^2) (R^2 + m^2) [(Q+R+P)^2 + m^2]} - \delta K_\phi^{(2)} P^2 + \dots
\end{align}
Again, after expanding in external momenta and applying the tensor reduction formulae, it is straightforward to express the result in terms of the following 2-loop bosonic sum-integrals:
\begin{align}
    \Gamma_{\phi_0 \phi_0}^{(2)} = 8 \lambda^2 |\mathbf{p}|^2 \left[ \frac{4}{d} \left( \hat{I}_{112}^{00} - \hat{I}_{113}^{02} - \hat{I}_{113}^{20} + 2 \hat{I}_{113}^{11} \right) - \hat{I}_{112}^{00} \right] - \delta K_\phi^{(2)} |\mathbf{p}|^2 + \dots
\end{align}

Lastly, the matching equation for $K_3$ is obtained from the condition that the renormalised 2-point correlator we just obtained is equal to the same in the 3D EFT.
Using the hard region expansion, and given that scaleless integrals vanish in dimensional regularisation, in the 3D EFT only tree-level diagrams remain. Thus:
\begin{equation}
    \Gamma_{\varphi_0 \varphi_0} = - K_3 |\mathbf{p}|^2 - m_3^2 - \delta m_3^2{}^{(2)}\,.
\end{equation}

Since only $m_3^2$ and $\lambda_3$ renormalise in the 3D EFT up to 2-loops, the matching equation for the kinetic terms is simply:
\begin{align}
    K_3 = &1 - 6 y^2 \left[ \frac{2}{d} \left(I_2^0 - I_3^2 \right) - I_2^0 \right] + \delta K_\phi^{(1)} \nonumber\\
    &- 8 \lambda^2  \left[ \frac{4}{d} \left( \hat{I}_{112}^{00} - \hat{I}_{113}^{02} - \hat{I}_{113}^{20} + 2 \hat{I}_{113}^{11} \right) - \hat{I}_{112}^{00} \right] + \delta K_\phi^{(2)}\,.
\end{align}
If we finally introduce the expression for the CT in Eq. \eqref{eq:4Dcounterterms} and solve the sum-integrals using the formulae in Appendix \ref{app:sumintegrals}, one  reproduces the result in Eq. \eqref{eq:matchingK3}.

Let us briefly note that it is also possible to compute the matching without including a CT for the wavefunction of the fields. In this method, one first canonically normalises the renormalised 4D Lagrangian, and then computes the matching equations with normalised CTs for each WC. For instance, let $F$ be a field with wavefunction CT denoted by $\delta K_F$, and $c_{n_F}$ and $\delta c_{n_F}$ be the WC of an operator with $n_F$ insertions of field $F$, and its CT, respectively. Then canonical normalisation amounts to taking
\begin{equation}
    c_{n_F} + \delta c_{n_F} \to c_{n_F} + \delta c'_{n_F} \equiv \frac{c_{n_F} + \delta c_{n_F}}{\left(1 + \delta K_F\right)^{n_F/2}}\,.
\end{equation}
Using these new $\delta c'_F$, one can derive a set of matching equations that will in principle be different from the ones computed when using the unnormalised $\delta c_F$. However, upon canonical normalisation of the matched WCs in the 3D EFT, both matching results can be seen to be exactly the same.

\section{Effective potential}
\label{app:effectivepotential}
The effective potential up to 2 loops within the 3D EFT can be expressed as follows:
\begin{align}
    V_\text{eff} &= m_3^2\varphi^\dagger\varphi+\lambda_3(\varphi^\dagger\varphi)^2+c_{\varphi^6}(\varphi^\dagger\varphi)^3+V_\text{eff}^\text{1-loop} + V_\text{eff}^\text{2-loop}\,, 
\end{align}
where the 1-loop parts reads
\begin{equation}
    V_\text{eff}^\text{1-loop} = -\frac{1}{12 \pi} m_\text{eff}^3\,,\quad m^2_\text{eff} = 2\left[m_3^2+4\lambda_3 \varphi^\dagger\varphi+9c_{\varphi^6}(\varphi^\dagger\varphi)^2\right]\,,
\end{equation}
and we compute the 2-loop contribution following Jackiw's background-field method~\cite{Jackiw:1974cv}. We restrict to this loop order because it suffices to remove essentially all renormalisation scale dependence and the finite pieces barely contribute to our numerical estimations.

To this aim, we write the Lagrangian in Eq.~\eqref{eq:3DLag} in terms of the real components of $\varphi=(\varphi_1+i\varphi_2)/\sqrt{2}$, and make the shift $\varphi_1\to\varphi_1+\tilde{\varphi}$. (This way we avoid mass mixing, while the full dependence on $\varphi$ can be later retrieved from $O(2)$ invariance upon replacing $\tilde{\varphi}\to\sqrt{2}\varphi$.)

\begin{figure}[t]

    \centering
    \includegraphics[width=0.19\textwidth]{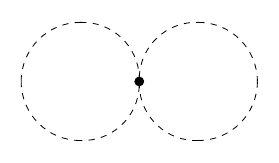}
    \includegraphics[width=0.12\textwidth]{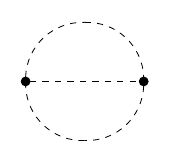}
    \caption{\it 2-loop vacuum diagrams.}\label{fig:vacuumdiagrams}
\end{figure}

Neglecting the dependence of the squared mass with $\tilde{\varphi}$ for simplicity, we obtain:
\begin{align}
    \mathcal{L}_{EFT} = \frac{1}{2}m_3^2\varphi_1^2+\kappa_1\varphi_1^3+\lambda_1\varphi_1^4+\frac{1}{2}m_3^2\varphi_2^2+\lambda_2\varphi_2^4+\kappa_{12}\varphi_1\varphi_2^2+\lambda_{12}\varphi_1^2\varphi_2^2+\cdots\,,
\end{align}
with
\begin{align}
    %
    %
    %
    \kappa_1 &=\lambda_3\tilde{\varphi}+\frac{5}{2}c_{\varphi^6}\tilde{\varphi}^3\,,\\
    \kappa_{12} &= \lambda_3\tilde{\varphi}+\frac{3}{2}c_{\varphi^6}\tilde{\varphi}^3\,,\\
    \lambda_1 &= \frac{1}{4}\lambda_3 + \frac{15}{8}c_{\varphi^6}\tilde{\varphi}^2\,,\\
    \lambda_2 &= \frac{1}{4}\lambda_3 + \frac{3}{8}c_{\varphi^6}\tilde{\varphi}^2\,,\\
    \lambda_{12} &= \frac{1}{2}\lambda_3 + \frac{9}{4}c_{\varphi^6}\tilde{\varphi}^2\,.
\end{align}
Thus, $V_\text{eff}^\text{2-loop}$ is given by the sum of the 2-loop vacuum diagrams computed with these field-dependent couplings; see Fig.~\ref{fig:vacuumdiagrams}.
We have:
\begin{align}
    V_\text{eff}^\text{2-loop} &= (3\lambda_1+3\lambda_2+\lambda_{12})\mc{I}_\text{bubble}^2(m_3)-(3\kappa_1^2+3\kappa_2^2+\kappa_{12}^2)\mc{I}_\text{sunset}(m_3)\nonumber\,,
\end{align}
where~\cite{Rajantie:1996np}
\begin{align}
    \mc{I}_\text{bubble}(m) = \frac{1}{(4\pi)^2}m^2\,, \quad \mc{I}_\text{sunset}(m) = \frac{1}{(4 \pi)^2} \left(\frac{1}{4 \epsilon} + \frac{1}{2} + \log{\frac{\mu}{3m}}\right)\,.
\end{align}

Altogether, and after removing $1/\epsilon$ poles and constant terms, we obtain:
\begin{align}\label{eq:effective_potential}
    V_\text{eff}^\text{2-loop} &= \frac{1}{4\pi^2} \bigg\lbrace\left[\frac{9}{2}m_3^2c_{\varphi^6}-\left(1+2\log{\frac{\mu}{3m_3}}\right)\lambda_3^2\right]\varphi^\dagger\varphi-9\left(1+2\log{\frac{\mu}{3m_3}}\right)c_{\varphi^6}\lambda_3(\varphi^\dagger\varphi)^2\bigg\rbrace\,.
\end{align}

\section{Real singlet scenario}
\label{app:othermodel}

We consider a second model consisting of a real scalar $\phi$ and a massless fermion $\psi$. The 4D Lagrangian in Minkowski space reads:
\begin{equation}
 \mathcal{L}_\text{4} = \frac{1}{2}(\partial\phi)^2 - \frac{1}{2}m^2\phi^2 - \kappa\phi^3 - \lambda\phi^4 + \overline{\psi} i\slashed{\partial} \psi - y \phi\overline{\psi}\psi\,.
 \label{eq:LagUV}
\end{equation}

As it was first studied in Ref.~\cite{Gould:2023jbz}, this model allows for non-radiatively-induced PTs, thanks to the presence of a non-$\mathbb{Z}_2$ symmetric cubic term for the scalar which contributes at tree-level to the formation of a barrier within the 3D effective potential. In the aforementioned work, PTs up to $\mathcal{O}(y^5)$ were studied, following the power counting prescription
\begin{equation}
    \frac{m^2}{T^2} \sim \frac{|\mathbf{p}|^2}{T^2} \sim \lambda \sim \frac{\kappa}{T} \sim y^2 \,,
\end{equation}
where we have omitted numerical factors. This amounts to 2-loop matching for the effective mass and 1-loop matching for the rest of the Wilson coefficients in a 3D EFT only including renormalisable interactions. In Ref.~\cite{Chala:2024xll}, leading corrections in the form of 1-loop dimension-6 and dimension-8 operators in the 3D EFT were also considered.

Here, we take this computation, including only fermion loops, to $O(y^6)$, except for the 3-loop contribution to the matching of the 3D effective mass. This is due to the current lack of knowledge about the evaluation of 3-loop, mass dimension 2, fermionic and mixed sum-integrals; see Appendix~\ref{app:sumintegrals} for further details. This leads to certain renormalisation-scale dependence in the mass term.

The scalar 3D EFT up to dimension-6 in Euclidean form reads~\cite{Chala:2024xll}:
\begin{equation}\label{eq:3deft_realscalar}
    \mathcal{L}_\text{EFT} = \frac{1}{2} K_3 (\partial\varphi)^2 + \frac{1}{2}m_3^2 \varphi^2 + \kappa_3 \varphi^3 + \lambda_3 \varphi^4 +c_{\varphi^6} \varphi^6 
    + r_{\partial^4 \varphi^2} \partial^2 \varphi \partial^2 \varphi 
    + r_{\partial^2 \varphi^4} \varphi^3 \partial^2 \varphi\,.
\end{equation}

Firstly, the 3D scalar field $\varphi$ is related to the zeroth Matsubara mode of $\phi$ through the following matching to the kinetic term~\footnote{Here we have corrected an error in the 1-loop matching of the kinetic term of the scalar published in Ref.~\cite{Chala:2024xll}, which yielded a $L_f$-independent contribution. The origin of the error is due to taking $d \to 3$ instead of $d \to 3 - 2\epsilon$ before expanding the result in $\epsilon$, as this $3-2\epsilon$ hits a pole in the expansion of the divergent sum-integral $\hat{I}_{2}^{0}$.}:
\begin{align}
    K_3 &= 1 + \frac{1}{8 \pi^2} y^2 L_f \nonumber\\
    &\hphantom{= 1 } - \frac{1}{3072 \pi^4} y^4 \left( 36 L_f^2 + 60 L_f + 93 - 64 \log 2 + 70 \zeta(3)  \right) \,.
\end{align}
We normalise canonically through $\varphi \to \varphi / \sqrt{K_3}$. Again, we use the same notation for normalised and unnormalised WCs, and the normalised matching equations for the first read as follows:
\begin{align}
    m_3^2 &= m^2 + y^2 \left( \frac{1}{6} T^2 - \frac{1}{8 \pi^2} m^2 L_f \right) \nonumber\\
    &\hphantom{=} + \frac{1}{768 \pi^2} y^2 \left\{ 8 T^2 \left( L_f - 8 \log 2 \right) + \frac{1}{4 \pi^2} m^2 \biggl[ 12 L_f^2 + 12 L_f \left( 1 + 48 \log 2 \right) \right. \nonumber\\
    &\hphantom{=} + 21 + 320 \log 2 + 70 \zeta(3) \biggr] \biggr\} + \frac{1}{18432 \pi^4} y^6 \biggl[ 12 L_f^2 + 12 L_f \left( 5 + 16 \log 2 \right) \nonumber\\
    &\hphantom{=} + 93 - 64 \log 2 + 70 \zeta(3) \biggr] \,,\\
    \kappa_3 &= \kappa \sqrt{T} - \frac{3}{16 \pi^2} \kappa y^2 \sqrt{T} L_f \nonumber\\
    &\hphantom{=} + \frac{1}{2048 \pi^4} \kappa y^4 \sqrt{T} \left( 96 L_f^2 + 60 L_f + 93 - 64 \log 2 + 70 \zeta(3) \right) \,,\\
    \lambda_3 &= \lambda T + \frac{1}{16 \pi^2} y^4 T L_f - \frac{1}{4 \pi^2} \lambda y^2 T L_f - \frac{1}{512 \pi^4} y^6 T \biggl[ 14 L_f^2 + 16 L_f + 24 + 7 \zeta(3) \biggr] \nonumber\\
    &\hphantom{=} + \frac{1}{1536 \pi^4} \lambda y^4 T \biggl[ 108 L_f^2 + 60 L_f + 93 - 64 \log 2 + 70 \zeta(3) \biggr] \,,\\
    c_{\varphi^6} &= - \frac{7 \zeta(3)}{192 \pi^4} y^6 \,,\\
    r_{\partial^4 \varphi^2} &= - \frac{7 \zeta(3)}{384 \pi^4 T^2} y^2 \,, \\
    r_{\partial^2 \varphi^4} &= \frac{35 \zeta(3)}{576 \pi^4 T} y^4 \,.
\end{align}

The UV CT Lagrangian reads:
\begin{equation}
 \mathcal{L}_{4, \rm ct} = \frac{1}{2} \delta K_\phi (\partial\phi)^2 - \frac{1}{2} \delta m ^2\phi^2 - \delta\kappa \phi^3 - \delta\lambda \phi^4 + \delta K_\psi \overline{\psi} i\slashed{\partial} \psi - \delta y \phi\overline{\psi}\psi\,.
 \label{eq:LagUV}
\end{equation}
We compute the CTs up to $\mathcal{O}(y^6)$, involving only fermion loops. To renormalise the matching equations we need the following up to 2-loops:
\begin{align}
    \delta K_\phi &= - \frac{y^2}{8 \pi^2 \epsilon} + \frac{y^4}{512 \pi^4} \left( \frac{5}{\epsilon} - \frac{6}{\epsilon^2} \right) \,,\\
    \delta m^{2} &= \frac{1}{128 \pi^4} y^4 m^2 \left( \frac{1}{\epsilon} - \frac{3}{\epsilon^2} \right) \,,\\
    \delta \lambda &= -\frac{y^4}{16 \pi^2 \epsilon} + \frac{y^6}{256 \pi^4} \left( \frac{4}{\epsilon} - \frac{3}{\epsilon^2} \right) \,,
\end{align}
while the rest are only needed up to 1-loop:
\begin{equation}
    \delta \kappa = 0\,, \quad \delta K_\psi = -\frac{y^2}{32 \pi^2 \epsilon} \,, \quad \delta y = \frac{y^3}{16 \pi^2 \epsilon} \,.
\end{equation}

On the other hand, the CT Lagrangian of the 3D Lagrangian reads, in Euclidean space:
\begin{align}\label{eq:3deft_realscalar}
    \mathcal{L}_\text{EFT, ct} &= \frac{1}{2} \delta K_\varphi (\partial\varphi)^2 + \frac{1}{2} \delta m_3{}^2 \varphi^2 + \delta \kappa_3 \varphi^3 + \delta \lambda_3 \varphi^4 + \delta c_{\varphi^6} \varphi^6 \nn\\
    &+ \delta r_{\partial^4 \varphi^2} \partial^2 \varphi \partial^2 \varphi 
    + \delta r_{\partial^2 \varphi^4} \varphi^3 \partial^2 \varphi\,,
\end{align}
and the only non-vanishing CT up to $\mathcal{O}(y^6)$ is:
\begin{align}
    \delta m_3{}^{2} = \frac{3}{2 \pi^2 \epsilon} \lambda_3^2 \,.
\end{align}

In the matching, we see that the UV CTs alone cancel all $\epsilon$ poles, and no temperature-dependent poles remain. Indeed, up to $\mathcal{O}(y^6)$, and without $\lambda$ and $\kappa$ insertions, this 3D CT does not contribute upon the substitution of the matching relations. This serves as a cross-check of our matching computation.

\bibliographystyle{style} 

\bibliography{refs}

\end{document}